\documentclass{LMCS}

\usepackage[dvips,all,graph]{xy}

\def\ifusecolor {\iftrue}
\usepackage{graphicx}

\def\SmallSize#1{\hbox to4pt{\hss$#1$}}

\def\SizeZero#1{\hbox to0pt{\hss$#1$}}

\mathchardef\pardot="0201
\mathchardef\seqdot="0201
\newxycolor{white}{White}
\ifusecolor
      \mathchardef\apardot="0201
      \catcode`\@=11
      \newxycolor@{seqcolor}{0 0 1}{rgb}{}{}
      \newxycolor@{antiseqcolor}{0.5 0.5 0}{rgb}{}{}
      \newxycolor@{antiparcolor}{0.5 0 0.5}{rgb}{}{}
      \newxycolor@{parcolor}{0 1 0}{rgb}{}{}
      \newxycolor@{aparcolor}{1 0 0}{rgb}{}{}
      \catcode`\@=12
   \else
      \mathchardef\apardot="0201
      \catcode`\@=11
      \newxycolor@{seqcolor}{0 0 0}{rgb}{}{}
      \newxycolor@{antiseqcolor}{0 0 0}{rgb}{}{}
      \newxycolor@{antiparcolor}{0 0 0}{rgb}{}{}
      \newxycolor@{parcolor}{0 0 0}{rgb}{}{}
      \newxycolor@{aparcolor}{0 0 0}{rgb}{}{}
      \catcode`\@=12
   \fi
   
\newbox\DerivOneBox
\newbox\DerivTwoBox
\newbox\DerivThreeBox
\newbox\DerivFourBox
\newdimen\DerivOneDimen
\newdimen\DerivTwoDimen
\newdimen\DerivThreeDimen
\newdimen\DerivFourDimen
\def\Derivationleaf #1#2#3#4#5{\global\setbox\derboxone=\hbox{\strut
                                    $\DerivationFactors{#1}{#2}{#3}{#4}{#5}11$}}%
\def\DerivationFactors #1#2#3#4#5#6#7{%
   \setbox\DerivOneBox=\hbox{$#1\strut$}%
      \DerivOneDimen=\wd\DerivOneBox\divide\DerivOneDimen by2
   \setbox\DerivThreeBox=\hbox{$#3\strut$}%
      \DerivThreeDimen=\wd\DerivThreeBox\divide\DerivThreeDimen by2
   \setbox\DerivTwoBox=\hbox{\box\DerivOneBox\hbox{$#2$}\box\DerivThreeBox}%
      \DerivTwoDimen=\wd\DerivTwoBox
   \setbox\DerivFourBox=\hbox{$#4\strut$}%
      \DerivFourDimen=\wd\DerivFourBox
   \ifdim\DerivFourDimen>\DerivTwoDimen
      \global\dercdim=\DerivFourDimen                
      \global\derldim=0pt                            
      \global\derrdim=0pt                            
      \advance\DerivFourDimen by-\DerivTwoDimen
      \divide \DerivFourDimen by2
      \advance\DerivTwoDimen  by-\DerivOneDimen
      \advance\DerivTwoDimen  by-\DerivThreeDimen
      \divide \DerivTwoDimen  by 2
   \else
      \global\dercdim=\DerivFourDimen                
      \DerivFourDimen=0pt
      \advance\DerivTwoDimen  by-\DerivOneDimen
      \advance\DerivTwoDimen  by-\DerivThreeDimen
      \global\derldim=\DerivTwoDimen
         \global\advance\derldim by-\dercdim
         \global\divide\derldim by2
         \global\advance\derldim by\DerivOneDimen    
      \global\derrdim=\DerivTwoDimen
         \global\advance\derrdim by-\dercdim
         \global\divide\derrdim by2
         \global\advance\derrdim by\DerivThreeDimen  
      \divide \DerivTwoDimen  by 2
   \fi
   \vbox{\offinterlineskip\hbox{\kern\DerivFourDimen\box\DerivTwoBox}%
         \hbox{\kern\DerivFourDimen\kern\DerivOneDimen
               \kern\DerivTwoDimen\kern-#6\DerivTwoDimen\hbox{$\xy
               0;<#6\DerivTwoDimen,0pt>:<0pt,#7\DerivTwoDimen>::
               (0,1);(2,1)**\crv{(1.25,1.1875)&(0.75,0.8125)};
               (1,0)**@{-};(0,1)**@{-};
               (1,0.625)*{\scriptstyle #5}
               \endxy$}}%
         \hbox{\kern\DerivFourDimen\kern\DerivOneDimen\kern\DerivTwoDimen
               \hbox to0pt{\hss\box\DerivFourBox\hss}%
               \kern\DerivFourDimen\kern\DerivOneDimen\kern\DerivTwoDimen}}}%

\def\atir  {{\ass\iss{\downarrow}}}%
\def\aatir {{\ass\iss{\uparrow  }}}%
\def\contr  {{\css\mkern-0mu\tss}}%
\def\BV {{\Bss\mkern-2mu\Vss}}%
\def\intr  {{\iss{\downarrow}}}%
\def\aintr {{\iss{\uparrow  }}}%
\def\seqr  {{\qss{\downarrow}}}%
\def\aseqr {{\qss{\uparrow  }}}%
\def\swir {\sss}
\def\SBV {{\Sss\mkern-1.5mu\Bss\mkern-2mu\Vss}}%
\def\sysS {{\script S}}%
\def\un {{\circ}}%
\def\unr {\un{\downarrow}}%

\def\sone       {{\sss_1}}%
\def\stwo       {{\sss_2}}%
\def\sthree     {{\sss_3}}%
\def\sfour      {{\sss_4}}%
\def\sfive      {{\sss_5}}%
\def\ssix       {{\sss_6}}%
\def\sseven     {{\sss_7}}%
\def\ssevenseq  {{\sss_7^{\seqrel}}}%
\def\ssevenpar  {{\sss_7^{\parrel}}}%
\def\ssevenapar {{\sss_7^{\aparrel}}}%

\def\occ {\mathop{\oss\mkern-.5mu\css\css}}%
\def\wb {\mathop{\wss}}

\def\merge {\mathbin{\bigdiamond}}%

\def\aparrel {\mathrel{\uparrow}}%
\def\aseqrel {\mathrel{\triangleright}}%
\let\colder=<

\def\grammareq {\mathrel{\raise.4pt\hbox{::}{=}}}%

\def\parrel {\mathrel{\downarrow}}%
\def\permover{\mathchoice
             {\displaystyle
              \mathrel{\lower.493\fontdimen5\textfont3
                       \hbox{$\lhook$}%
                       \mkern-3mu{\rightarrow}}}%
             {\textstyle
              \mathrel{\lower.493\fontdimen5\textfont3
                       \hbox{$\lhook$}%
                       \mkern-3mu{\rightarrow}}}%
             {\scriptstyle
              \mathrel{\lower.54\fontdimen5\scriptfont3
                       \hbox{$\scriptstyle\lhook$}%
                       \mkern-3.2mu{\rightarrow}}}%
             {\scriptscriptstyle
              \mathrel{\lower.55\fontdimen5\scriptscriptfont3
                       \hbox{$\scriptscriptstyle\lhook$}%
                       \mkern-3.2mu{\rightarrow}}}%
             }%
\def\seqrel {\mathrel{\triangleleft}}%
\let\neg=\bar

\input rotate.tex

\def\hexnumber #1{\ifcase #10\or 1\or 2\or 3\or 4\or 5\or 6\or 7\or 8\or
   9\or A\or B\or C\or D\or E\or F\fi}%

\newfam\cmsmrfam
   \font\twelvesmr=cmsmr10 at 12pt
   \font\tensmr=cmsmr10
   \font\ninesmr=cmsmr10 at 9pt
   \font\eightsmr=cmsmr10 at 8pt
   \font\sevensmr=cmsmr7
   \font\sixsmr=cmsmr7 at 6pt
   \font\fivesmr=cmsmr5
   \textfont\cmsmrfam=\tensmr \scriptfont\cmsmrfam=\sevensmr
      \scriptscriptfont\cmsmrfam=\fivesmr

   \font\ninesmiu=cmsmiu10 at 9pt
   \font\eightsmiu=cmsmiu10 at 8pt

\newfam\eufmfam
   \font\teneufm=eufm10
   \font\nineeufm=eufm10 at 9pt
   \font\eighteufm=eufm10 at 8pt
   \font\seveneufm=eufm7
   \font\sixeufm=eufm7 at 6pt
   \font\fiveeufm=eufm5
   \textfont\eufmfam=\teneufm \scriptfont\eufmfam=\seveneufm
      \scriptscriptfont\eufmfam=\fiveeufm

\newfam\eurmfam
   \font\teneurm=eurm10
   \font\nineeurm=eurm10 at 9pt
   \font\eighteurm=eurm10 at 8pt
   \font\seveneurm=eurm7
   \font\sixeurm=eurm7 at 6pt
   \font\fiveeurm=eurm5
   \textfont\eurmfam=\teneurm \scriptfont\eurmfam=\seveneurm
      \scriptscriptfont\eurmfam=\fiveeurm

\newfam\eusmfam
   \font\teneusm=eusm10
   \font\nineeusm=eusm10 at 9pt
   \font\eighteusm=eusm10 at 8pt
   \font\seveneusm=eusm7
   \font\sixeusm=eusm7 at 6pt
   \font\fiveeusm=eusm5
   \textfont\eusmfam=\teneusm \scriptfont\eusmfam=\seveneusm
      \scriptscriptfont\eusmfam=\fiveeusm

\newfam\msamfam
   \font\tenmsam=msam10
   \font\ninemsam=msam10 at 9pt
   \font\eightmsam=msam10 at 8pt
   \font\sevenmsam=msam7
   \font\sixmsam=msam7 at 6pt
   \font\fivemsam=msam5
   \textfont\msamfam=\tenmsam \scriptfont\msamfam=\sevenmsam
      \scriptscriptfont\msamfam=\fivemsam

\newfam\msbmfam
   \font\tenmsbm=msbm10
   \font\ninemsbm=msbm10 at 9pt
   \font\eightmsbm=msbm10 at 8pt
   \font\sevenmsbm=msbm7
   \font\sixmsbm=msbm7 at 6pt
   \font\fivemsbm=msbm5
   \textfont\msbmfam=\tenmsbm \scriptfont\msbmfam=\sevenmsbm
      \scriptscriptfont\msbmfam=\fivemsbm

\newfam\scriptfam
      \font\tenfs=rsfs10 \skewchar\tenfs='177
      \font\ninefs=rsfs10 at 9pt \skewchar\ninefs='177
      \font\eightfs=rsfs10 at 8pt \skewchar\eightfs='177
      \font\sevenfs=rsfs7 \skewchar\sevenfs='177
      \font\sixfs=rsfs7 at 6pt \skewchar\sixfs='177
      \font\fivefs=rsfs5 \skewchar\fivefs='177
      \textfont\scriptfam=\tenfs \scriptfont\scriptfam=\sevenfs
         \scriptscriptfont\scriptfam=\fivefs

\newfam\stmaryrdfam
   \font\tenstmaryrd=stmary10
   \font\ninestmaryrd=stmary10 at 9pt
   \font\eightstmaryrd=stmary10 at 8pt
   \font\sevenstmaryrd=stmary7
   \font\sixstmaryrd=stmary7 at 6pt
   \font\fivestmaryrd=stmary5
   \textfont\stmaryrdfam=\tenstmaryrd \scriptfont\stmaryrdfam=\sevenstmaryrd
      \scriptscriptfont\stmaryrdfam=\fivestmaryrd

\def\eightpointlogic{%
   \textfont\eurmfam=\eighteurm \scriptfont\eurmfam=\sixeurm
      \scriptscriptfont\eurmfam=\fiveeurm
   \textfont\cmsmrfam=\eightsmr \scriptfont\cmsmrfam=\sixsmr
      \scriptscriptfont\cmsmrfam=\fivesmr
   \textfont\cmsmiufam=\eightsmiu \scriptfont\cmsmiufam=\sixsmiu
      \scriptscriptfont\cmsmiufam=\fivesmiu
   \textfont\eufmfam=\eighteufm \scriptfont\eufmfam=\sixeufm
      \scriptscriptfont\eufmfam=\fiveeufm
   \textfont\eusmfam=\eighteusm \scriptfont\eusmfam=\sixeusm
      \scriptscriptfont\eusmfam=\fiveeusm
   \textfont\msamfam=\eightmsam \scriptfont\msamfam=\sixmsam
      \scriptscriptfont\msamfam=\fivemsam
   \textfont\msbmfam=\eightmsbm \scriptfont\msbmfam=\sixmsbm
      \scriptscriptfont\msbmfam=\fivemsbm
   \ifscriptfamrsfs
      \textfont\scriptfam=\eightfs \scriptfont\scriptfam=\sixfs
         \scriptscriptfont\scriptfam=\fivefs
      \else
      \textfont\scriptfam=\eightmptwo \scriptfont\scriptfam=\sixmptwo
         \scriptscriptfont\scriptfam=\fivemptwo
      \fi
   \textfont\stmaryrdfam=\eightstmaryrd \scriptfont\stmaryrdfam=\sixstmaryrd
      \scriptscriptfont\stmaryrdfam=\fivestmaryrd}%
\let\oldeightpoint=\eightpoint
\def\eightpoint {\oldeightpoint\eightpointlogic}%

\def\ninepointlogic{%
   \textfont\eurmfam=\nineeurm \scriptfont\eurmfam=\sixeurm
      \scriptscriptfont\eurmfam=\fiveeurm
   \textfont\cmsmrfam=\ninesmr \scriptfont\cmsmrfam=\sixsmr
      \scriptscriptfont\cmsmrfam=\fivesmr
   \textfont\cmsmiufam=\ninesmiu \scriptfont\cmsmiufam=\sixsmiu
      \scriptscriptfont\cmsmiufam=\fivesmiu
   \textfont\eufmfam=\nineeufm \scriptfont\eufmfam=\sixeufm
      \scriptscriptfont\eufmfam=\fiveeufm
   \textfont\eusmfam=\nineeusm \scriptfont\eusmfam=\sixeusm
      \scriptscriptfont\eusmfam=\fiveeusm
   \textfont\msamfam=\ninemsam \scriptfont\msamfam=\sixmsam
      \scriptscriptfont\msamfam=\fivemsam
   \textfont\msbmfam=\ninemsbm \scriptfont\msbmfam=\sixmsbm
      \scriptscriptfont\msbmfam=\fivemsbm
   \ifscriptfamrsfs
      \textfont\scriptfam=\ninefs \scriptfont\scriptfam=\sixfs
         \scriptscriptfont\scriptfam=\fivefs
      \else
      \textfont\scriptfam=\ninemptwo \scriptfont\scriptfam=\sixmptwo
         \scriptscriptfont\scriptfam=\fivemptwo
      \fi
   \textfont\stmaryrdfam=\ninestmaryrd \scriptfont\stmaryrdfam=\sixstmaryrd
      \scriptscriptfont\stmaryrdfam=\fivestmaryrd}%
\let\oldninepoint=\ninepoint
\def\ninepoint {\oldninepoint\ninepointlogic}%

\def\twelvepointlogic{%
   \textfont\cmsmrfam=\twelvesmr \scriptfont\cmsmrfam=\eightsmr
      \scriptscriptfont\cmsmrfam=\sixsmr
      }%
\let\oldtwelvepoint=\twelvepoint
\def\twelvepoint {\oldtwelvepoint\twelvepointlogic}%

\def\script {\fam\scriptfam}

\catcode`\!=\active
   \edef!{\hexnumber\eurmfam}%
      \mathchardef\Deltabb="0!01
      \mathchardef\Pibb="0!05
      \mathchardef\Sigmabb="0!06
      \mathchardef\gammabb="0!0D
      \mathchardef\deltabb="0!0E
      \mathchardef\kappabb="0!14
      \mathchardef\pibb="0!19
      \mathchardef\psibb="0!20
      \mathchardef\rhobb="0!1A
      \mathchardef\sigmabb="0!1B
      \mathchardef\Abb="0!41
      \mathchardef\Bbb="0!42
      \mathchardef\Cbb="0!43
      \mathchardef\Dbb="0!44
      \mathchardef\Ebb="0!45
      \mathchardef\Fbb="0!46 
      \mathchardef\Gbb="0!47
      \mathchardef\Hbb="0!48
      \mathchardef\Ibb="0!49
      \mathchardef\Jbb="0!4A
      \mathchardef\Kbb="0!4B
      \mathchardef\Lbb="0!4C
      \mathchardef\Mbb="0!4D
      \mathchardef\Nbb="0!4E
      \mathchardef\Obb="0!4F
      \mathchardef\Pbb="0!50
      \mathchardef\Qbb="0!51
      \mathchardef\Rbb="0!52
      \mathchardef\Sbb="0!53
      \mathchardef\Tbb="0!54
      \mathchardef\Ubb="0!55
      \mathchardef\Vbb="0!56
      \mathchardef\Wbb="0!57
      \mathchardef\Xbb="0!58
      \mathchardef\Ybb="0!59
      \mathchardef\Zbb="0!5A
      \mathchardef\abb="0!61
      \mathchardef\bbb="0!62
      \mathchardef\cbb="0!63
      \mathchardef\dbb="0!64
      \mathchardef\ebb="0!65
      \mathchardef\fbb="0!66
      \mathchardef\gbb="0!67
      \mathchardef\hbb="0!68
      \mathchardef\ibb="0!69
      \mathchardef\jbb="0!6A
      \mathchardef\kbb="0!6B
      \mathchardef\lbb="0!6C
      \mathchardef\mbb="0!6D
      \mathchardef\nbb="0!6E
      \mathchardef\obb="0!6F
      \mathchardef\pbb="0!70
      \mathchardef\qbb="0!71
      \mathchardef\rbb="0!72
      \mathchardef\sbb="0!73
      \mathchardef\tbb="0!74
      \mathchardef\ubb="0!75
      \mathchardef\vbb="0!76
      \mathchardef\wbb="0!77
      \mathchardef\xbb="0!78
      \mathchardef\ybb="0!79
      \mathchardef\zbb="0!7A
   \edef!{\hexnumber\cmsmrfam}%
      \mathchardef\Ass="0!41
      \mathchardef\Bss="0!42
      \mathchardef\Css="0!43
      \mathchardef\Dss="0!44
      \mathchardef\Ess="0!45
      \mathchardef\Fss="0!46 
      \mathchardef\Gss="0!47
      \mathchardef\Hss="0!48
      \mathchardef\Iss="0!49
      \mathchardef\Jss="0!4A
      \mathchardef\Kss="0!4B
      \mathchardef\Lss="0!4C
      \mathchardef\Mss="0!4D
      \mathchardef\Nss="0!4E
      \mathchardef\Oss="0!4F
      \mathchardef\Pss="0!50
      \mathchardef\Qss="0!51
      \mathchardef\Rss="0!52
      \mathchardef\Sss="0!53
      \mathchardef\Tss="0!54
      \mathchardef\Uss="0!55
      \mathchardef\Vss="0!56
      \mathchardef\Wss="0!57
      \mathchardef\Xss="0!58
      \mathchardef\Yss="0!59
      \mathchardef\Zss="0!5A
      \mathchardef\ass="0!61
      \mathchardef\bss="0!62
      \mathchardef\css="0!63
      \mathchardef\dss="0!64
      \mathchardef\ess="0!65
      \mathchardef\fss="0!66
      \mathchardef\gss="0!67
      \mathchardef\hhss="0!68    
      \mathchardef\iss="0!69
      \mathchardef\jss="0!6A
      \mathchardef\kss="0!6B
      \mathchardef\lss="0!6C
      \mathchardef\mss="0!6D
      \mathchardef\nss="0!6E
      \mathchardef\oss="0!6F
      \mathchardef\pss="0!70
      \mathchardef\qss="0!71
      \mathchardef\rss="0!72
      \mathchardef\sss="0!73
      \mathchardef\tss="0!74
      \mathchardef\uss="0!75
      \mathchardef\vvss="0!76    
      \mathchardef\wss="0!77
      \mathchardef\xss="0!78
      \mathchardef\yss="0!79
      \mathchardef\zss="0!7A
      \mathchardef\bang="0!21
      \mathchardef\loc="0!21
      \mathchardef\lone="0!31
      \mathchardef\ltwo="0!32   
      \mathchardef\lwn="0!3F
      \mathchardef\lzero="0!30
      \mathchardef\Gammass="0!00
      \mathchardef\Deltass="0!01
      \mathchardef\deltass="0!0E
      \mathchardef\gitss="0!67
   \edef!{\hexnumber\msamfam}%
      \mathchardef\bigdiamond="0!06
      \mathchardef\triangleup="0!4D
      \mathchardef\triangledown="0!4F
   \edef!{\hexnumber\msbmfam}%
      \mathchardef\emptyset="0!3F
   \edef!{\hexnumber\scriptfam}%
         \mathchardef\Asc="0!41
         \mathchardef\Bsc="0!42
         \mathchardef\Csc="0!43
         \mathchardef\Dsc="0!44
         \mathchardef\Esc="0!45
         \mathchardef\Fsc="0!46 
         \mathchardef\Gsc="0!47
         \mathchardef\Hsc="0!48
         \mathchardef\Isc="0!49
         \mathchardef\Jsc="0!4A
         \mathchardef\Ksc="0!4B
         \mathchardef\Lsc="0!4C
         \mathchardef\Msc="0!4D
         \mathchardef\Nsc="0!4E
         \mathchardef\Osc="0!4F
         \mathchardef\Psc="0!50
         \mathchardef\Qsc="0!51
         \mathchardef\Rsc="0!52
         \mathchardef\Ssc="0!53
         \mathchardef\Tsc="0!54
         \mathchardef\Usc="0!55
         \mathchardef\Vsc="0!56
         \mathchardef\Wsc="0!57
         \mathchardef\Xsc="0!58
         \mathchardef\Ysc="0!59
         \mathchardef\Zsc="0!5A
   \edef!{\hexnumber\cmsmrfam}%
\catcode`\!=12
\mathchardef\contr="013E
\newbox\Nablassbox
\def\Nablass {{\mathchoice
      {\setbox\Nablassbox=\hbox{$\Deltass$}%
         \setbox\Nablassbox=\hbox{\rotu\Nablassbox}%
         \box\Nablassbox}%
      {\setbox\Nablassbox=\hbox{$\Deltass$}%
         \setbox\Nablassbox=\hbox{\rotu\Nablassbox}%
         \box\Nablassbox}%
      {\setbox\Nablassbox=\hbox{$\scriptstyle\Deltass$}%
         \setbox\Nablassbox=\hbox{\rotu\Nablassbox}%
         \box\Nablassbox}%
      {\setbox\Nablassbox=\hbox{$\scriptscriptstyle\Deltass$}%
         \setbox\Nablassbox=\hbox{\rotu\Nablassbox}%
         \box\Nablassbox}}}%

\mathchardef\weak="013C
\mathchardef\impl="221B
\def\limp {\mathchoice
    {\mathbin{{-}\mkern-3.2mu{\circ}}}%
    {\mathbin{{-}\mkern-3.2mu{\circ}}}%
    {\mathbin{{-}\mkern-3.5mu{\circ}}}%
    {\mathbin{{-}\mkern-3.7mu{\circ}}}}%
\catcode`\!=\active
   \edef!{\hexnumber\stmaryrdfam}%
   \mathchardef\squarebox="2!1F
   \mathchardef\lpar="2!4F
   \mathchardef\lplus="2!16
   \mathchardef\lmix="2!22
   \mathchardef\lprec="2!34
   \mathchardef\ltens="2!0F
   \mathchardef\lwith="2!4E
   \mathchardef\merge="2!05
   \edef!{\hexnumber\msamfam}%
   \mathchardef\limpalt="2!28
   \catcode`\!=12

\catcode`\!=\active
   \edef!{\hexnumber\msamfam}%
   \mathchardef\ge="3!3E
   \mathchardef\le="3!36
   \mathchardef\gex="3!3C
   \mathchardef\lex="3!34
   \edef!{\hexnumber\stmaryrdfam}%
   \mathchardef\bigcross="3!22
   \mathchardef\arrowequiv="3!2D
   \mathchardef\inplus="3!41
   \mathchardef\msin="3!41
   \mathchardef\subsetplus="3!44
   \mathchardef\supsetplus="3!45
   \mathchardef\subseteqplus="3!46
   \mathchardef\supseteqplus="3!47
   \catcode`\!=12

\let\turnstile=\vdash

\catcode`\!=\active
   \edef!{\hexnumber\stmaryrdfam}%
   \mathchardef\lbbrack="4!4A
   \mathchardef\rbbrack="5!4B
   \mathchardef\lbpar="4!4C
   \mathchardef\rbpar="5!4D
      \mathchardef\lstrange="4!2A
      \mathchardef\rstrange="5!2B
      \mathchardef\lstrange="4!48
      \mathchardef\rstrange="5!49
   \catcode`\!=12

\def\xyvdots {\raise6pt\hbox{$\vdots$}}%

\newdimen\dercldim                                
\newdimen\derccdim                                
\newdimen\dercrdim                                
\newdimen\derldim                                 
\newdimen\dercdim                                 
\newdimen\derrdim                                 
\newdimen\derdim                                  
\newdimen\derdldim                                
\newdimen\derdrdim                                
\newbox\derboxone                                 
\newbox\derboxtwo                                 
\newbox\derboxthree                               
\newbox\derboxfour                                
\newdimen\derquad\derquad=\fontdimen6\textfont2
\newdimen\deropen\deropen=\fontdimen5\textfont2\divide\deropen by3
\def\leaf #1{\global\setbox\derboxone=\hbox{\strut$#1$}%
   \global\derldim=0pt                            
   \global\dercdim=\wd\derboxone                  
   \global\derrdim=0pt                            
   }%
\def\rootaux #1#2#3{\setbox\derboxtwo=\hbox{\unhbox\derboxone}%
   \setbox\derboxthree=\hbox 
      {$\smash{\lower\fontdimen22\textfont2\hbox{$#1$}}$}%
   \setbox\derboxfour=\hbox 
      {$\smash{\lower\fontdimen22\textfont2\hbox{$#2$}}$}%
   \leaf{#3}
   \derdim=\dercdim\advance\derdim by-\derccdim\divide\derdim by2 
   \global\derldim=\dercldim\global\advance\derldim by-\derdim
   \global\derrdim=\dercrdim\global\advance\derrdim by-\derdim
   \deropen=\fontdimen5\textfont2\divide\deropen by3
   \setbox\derboxone=\hbox{\vbox{\offinterlineskip
         \hbox{\ifdim\derldim<0pt\kern-\derldim\fi
               \box\derboxtwo
               \ifdim\derrdim<0pt\kern-\derrdim\fi}%
         \kern\deropen
         \hbox{\ifdim\dercldim>\derldim
                  \ifdim\derldim>0pt\kern\derldim\fi
                  \else\kern\dercldim\fi
               \hbox to0pt{\hss\copy\derboxthree}%
               \vbox{\ifdim\derccdim>\dercdim\hsize=\derccdim
                                        \else\hsize=\dercdim \fi
                    \hrule height.2pt depth.2pt width\hsize}%
               \hbox to0pt{\copy\derboxfour\hss}%
               \ifdim\dercrdim>\derrdim
                  \ifdim\derrdim>0pt\kern\derrdim\fi
                  \else\kern\dercrdim\fi}%
         \kern\deropen
         \hbox{\ifdim\derldim>0pt\kern\derldim\fi
               \box\derboxone
               \ifdim\derrdim>0pt\kern\derrdim\fi}}}%
   \ifdim\derldim<0pt\global\derldim=0pt\fi       
   \ifdim\derrdim<0pt\global\derrdim=0pt\fi       
   \derdldim=\wd\derboxthree\advance\derdldim by-\dercldim
   \derdrdim=\wd\derboxfour \advance\derdrdim by-\dercrdim
   \ifdim\derdim<0pt
      \ifdim\derdldim<0pt
         \derdldim=0pt                            
      \fi
      \ifdim\derdrdim<0pt
         \derdrdim=0pt                            
      \fi
   \else
      \ifdim\derldim>0pt
         \ifdim\derdldim>-\derdim
            \advance\derdldim by\derdim           
         \else                                            
            \derdldim=0pt                         
         \fi                                      
      \else
         \advance\derdldim by\dercldim            
      \fi
      \ifdim\derrdim>0pt
         \ifdim\derdrdim>-\derdim
            \advance\derdrdim by\derdim           
         \else                                            
            \derdrdim=0pt                         
         \fi                                      
      \else
         \advance\derdrdim by\dercrdim            
      \fi
   \fi
   \global\setbox\derboxone=\hbox
      {\kern\derdldim\unhbox\derboxone\kern\derdrdim}%
   \global\advance\derldim by\derdldim            
   \global\advance\derrdim by\derdrdim            
   }%
\def\rootr #1#2#3#4{{#4}%
   \dercldim=\derldim
   \derccdim=\dercdim
   \dercrdim=\derrdim
   \rootaux{#1}{#2}{#3}}%
\def\rrootr #1#2#3#4#5{\derquad=\fontdimen6\textfont2
   {#4}%
           \dercldim  =\derldim
   \setbox\derboxtwo=\hbox{\unhbox\derboxone\kern\derquad}%
           \derccdim  =\dercdim
   \advance\derccdim by\derrdim
   \advance\derccdim by\derquad
   {#5}%
   \setbox\derboxone=\hbox{\unhbox\derboxtwo\unhbox\derboxone}%
   \advance\derccdim by\derldim
   \advance\derccdim by\dercdim
           \dercrdim  =\derrdim
   \rootaux{#1}{#2}{#3}}%
\def\rrrootr #1#2#3#4#5#6{\derquad=\fontdimen6\textfont2
   {#4}%
           \dercldim  =\derldim
   \setbox\derboxtwo=\hbox{\unhbox\derboxone\kern\derquad}%
           \derccdim  =\dercdim
   \advance\derccdim by\derrdim
   \advance\derccdim by\derquad
   {#5}%
   \setbox\derboxtwo=\hbox{\unhbox\derboxtwo\unhbox\derboxone\kern\derquad}%
   \advance\derccdim by\derldim
   \advance\derccdim by\dercdim
   \advance\derccdim by\derrdim
   \advance\derccdim by\derquad
   {#6}%
   \setbox\derboxone=\hbox{\unhbox\derboxtwo\unhbox\derboxone}%
   \advance\derccdim by\derldim
   \advance\derccdim by\dercdim
           \dercrdim  =\derrdim
   \rootaux{#1}{#2}{#3}}%
\def\root       #1#2#3{\rootr  {#1\;}{}{#2}{#3}}%
\def\rroot    #1#2#3#4{\rrootr {#1\;}{}{#2}{#3}{#4}}%
\def\deraux {\derldim=0pt\dercdim=0pt\derrdim=0pt}%
\def\der       #1#2#3{\deraux\root  {#1}{#2}{#3}        \box\derboxone}%
\def\dder    #1#2#3#4{\deraux\rroot {#1}{#2}{#3}{#4}    \box\derboxone}%
\def\dernote       #1#2#3#4{\deraux\rootr  {#1\;}{\;#2}{#3}{#4}\box\derboxone}%
\def\inf       #1#2#3{\der  {#1}{#2}{\leaf{#3}}}%
\def\iinf    #1#2#3#4{\dder {#1}{#2}{\leaf{#3}}{\leaf{#4}}}%
\def\infnote       #1#2#3#4{\dernote  {#1}{#4}{#2}{\leaf{#3}}}%
\newbox\derskelboxone
\newbox\derskelboxtwo
\newbox\derskelboxthree
\newbox\derskelboxfour
\newdimen\derskeldimenone
\newdimen\derskeldimentwo
\newdimen\derskeldimenthree
\newdimen\derskeldimenfour
\newdimen\derskeldimenfive
\newdimen\derskeldimensix
\newdimen\derskeldimenseven
\newdimen\derskeldimeneight
\def\derskel #1#2#3#4{%
   \setbox\derskelboxone=\hbox{$#1$\strut}%
   \derskeldimenone=\ht\derskelboxone
   \advance\derskeldimenone by\dp\derskelboxone
   \derskeldimentwo=\wd\derskelboxone
   \divide\derskeldimentwo by2
   \setbox\derskelboxone=\hbox to0pt{%
      \hss\raise\dp\derskelboxone\box\derskelboxone\hss}%
   \ht\derskelboxone=0pt
   \dp\derskelboxone=0pt
   \setbox\derskelboxtwo=\hbox{$#3$\strut}%
   \derskeldimenthree=\ht\derskelboxtwo
   \advance\derskeldimenthree by\dp\derskelboxtwo
   \derskeldimenfour=\wd\derskelboxtwo
   \divide\derskeldimenfour by2
   \setbox\derskelboxtwo=\hbox to0pt{%
      \hss\raise\dp\derskelboxtwo\box\derskelboxtwo\hss}%
   \ht\derskelboxtwo=0pt
   \dp\derskelboxtwo=0pt
   \ifdim\derskeldimenone>\derskeldimenthree
      \else\derskeldimenone=\derskeldimenthree\fi
   \setbox\derskelboxthree=\hbox{$#4$\strut}%
   \derskeldimenfive=\ht\derskelboxthree
   \advance\derskeldimenfive by\dp\derskelboxthree
   \derskeldimensix=\wd\derskelboxthree
   \divide\derskeldimensix by2
   \setbox\derskelboxthree=\hbox to0pt{%
      \hss\lower\ht\derskelboxthree\box\derskelboxthree\hss}%
   \ht\derskelboxthree=0pt
   \dp\derskelboxthree=0pt
   \setbox\derskelboxfour=\hbox{$#2$\strut}%
   \derskeldimenseven=\ht\derskelboxfour
   \advance\derskeldimenseven by\dp\derskelboxfour
   \derskeldimeneight=\wd\derskelboxfour
   \divide\derskeldimeneight by2
   \setbox\derskelboxfour=\hbox to0pt{%
      \hss\raise\dp\derskelboxfour\box\derskelboxfour\hss}%
   \ht\derskelboxfour=0pt
   \dp\derskelboxfour=0pt
   \ifdim\derskeldimenone>\derskeldimenseven
      \else\derskeldimenone=\derskeldimenseven\fi
   \derskeldimenthree=\derskeldimentwo
   \advance\derskeldimenthree by2\derskeldimeneight
   \advance\derskeldimenthree by\derskeldimenfour
   \advance\derskeldimenthree by2em
   \divide\derskeldimenthree by2
   \advance\derskeldimensix by-\derskeldimenthree
   \derskeldimenseven=\derskeldimensix
   \advance\derskeldimensix by-\derskeldimentwo
   \advance\derskeldimenseven by-\derskeldimenfour
   \ifdim\derskeldimensix>0pt
      \else\derskeldimensix=0pt\fi
   \ifdim\derskeldimenseven>0pt
      \else\derskeldimenseven=0pt\fi
   \vbox{\kern\derskeldimenone\hbox{\kern\derskeldimensix
         \kern\derskeldimentwo
         \xy
         <-\derskeldimenthree,\derskeldimenthree>="here"
            *{\box\derskelboxone}**\dir{-};
         "here"+<\derskeldimentwo,0pt>="here"**\dir{-};
         "here"+<1em,0pt>="here"**\dir{-};
         "here"+<\derskeldimeneight,0pt>="here"
            *{\box\derskelboxfour}**\dir{-};
         "here"+<\derskeldimeneight,0pt>="here"**\dir{-};
         "here"+<1em,0pt>="here"**\dir{-};
         "here"+<\derskeldimenfour,0pt>*{\box\derskelboxtwo}**\dir{-};
         0*{\box\derskelboxthree}**\dir{-};
         <-\derskeldimenthree,\derskeldimenthree>**\dir{-}
         \endxy
         \kern\derskeldimenfour\kern\derskeldimenseven}%
      \kern\derskeldimenfive}}%

\def\boxtherule #1#2#3#4{\hbox{%
      \vtop{\kern0pt\hbox to0pt{\hss\structcolor{#1}\strut\enspace}}%
      \vtop{\kern0pt\structcolor{\boxit{\hbox to#2{\hfil\vbox to#3{%
                  \vfil\hbox to0pt{\hss\Black{$#4$}\hss}\vfil}\hfil}}}}}}%
\def\lowerdertobase #1{\lower
                       \ifsmallprint
                          6.89164pt
                       \else\ifverysmallprint
                          6.14815pt
                       \else
                          7.63518pt
                       \fi\fi\hbox{$#1$}}%

\def\InvisibleMark {\White{\vbox to0pt{\vss
   \hbox to0pt{\hss\vrule height1sp depth0pt width1sp}}}}%
\def\InvisibleMarkDown  #1{\kern-.#1pc\vbox to0pt{\kern.#1pc\InvisibleMark\vss}}%
\def\InvisibleMarkDDown #1{\kern-#1pc\vbox  to0pt{\kern#1pc\InvisibleMark \vss}}%
\def\InvisibleMarkUp    #1{\vbox to0pt{\vss\InvisibleMark\kern.#1pc}\kern-.#1pc}%
\def\InvisibleMarkUUp   #1{\vbox to0pt{\vss\InvisibleMark\kern#1pc}\kern-#1pc}%

\def\conldel {\{}%
\def\conrdel {\}}%
\def\lrgldel {\mathchoice{(}{(}{\langle}{\langle}}%
\def\lrgrdel {\mathchoice{)}{)}{\rangle}{\rangle}}%
\def\aprldel {\mathchoice
   {\mathopen {\setbox0=\hbox{$\displaystyle     \lrgldel$}\hbox to\wd0
                        {\hfil$\displaystyle     (       $\hfil}}}%
   {\mathopen {\setbox0=\hbox{$\textstyle        \lrgldel$}\hbox to\wd0
                        {\hfil$\textstyle        (        $\hfil}}}%
   {\mathopen {\setbox0=\hbox{$\scriptstyle      \lrgldel$}\hbox to\wd0
                        {\hfil$\scriptstyle      (        $\hfil}}}%
   {\mathopen {\setbox0=\hbox{$\scriptscriptstyle\lrgldel$}\hbox to\wd0
                        {\hfil$\scriptscriptstyle(        $\hfil}}}}%
\def\aprrdel {\mathchoice
   {\mathclose{\setbox0=\hbox{$\displaystyle     \lrgrdel$}\hbox to\wd0
                        {\hfil$\displaystyle     )       $\hfil}}}%
   {\mathclose{\setbox0=\hbox{$\textstyle        \lrgrdel$}\hbox to\wd0
                        {\hfil$\textstyle        )        $\hfil}}}%
   {\mathclose{\setbox0=\hbox{$\scriptstyle      \lrgrdel$}\hbox to\wd0
                        {\hfil$\scriptstyle      )        $\hfil}}}%
   {\mathclose{\setbox0=\hbox{$\scriptscriptstyle\lrgrdel$}\hbox to\wd0
                        {\hfil$\scriptscriptstyle)        $\hfil}}}}%
\def\seqldel {\mathchoice
   {\mathopen {\setbox0=\hbox{$\displaystyle     \lrgldel$}\hbox to\wd0
                        {\hfil$\displaystyle     \langle  $\hfil}}}%
   {\mathopen {\setbox0=\hbox{$\textstyle        \lrgldel$}\hbox to\wd0
                        {\hfil$\textstyle        \langle  $\hfil}}}%
   {\mathopen {\setbox0=\hbox{$\scriptstyle      \lrgldel$}\hbox to\wd0
                        {\hfil$\scriptstyle      \langle  $\hfil}}}%
   {\mathopen {\setbox0=\hbox{$\scriptscriptstyle\lrgldel$}\hbox to\wd0
                        {\hfil$\scriptscriptstyle\langle  $\hfil}}}}%
\def\seqrdel {\mathchoice
   {\mathclose{\setbox0=\hbox{$\displaystyle     \lrgrdel$}\hbox to\wd0
                        {\hfil$\displaystyle     \rangle  $\hfil}}}%
   {\mathclose{\setbox0=\hbox{$\textstyle        \lrgrdel$}\hbox to\wd0
                        {\hfil$\textstyle        \rangle  $\hfil}}}%
   {\mathclose{\setbox0=\hbox{$\scriptstyle      \lrgrdel$}\hbox to\wd0
                        {\hfil$\scriptstyle      \rangle  $\hfil}}}%
   {\mathclose{\setbox0=\hbox{$\scriptscriptstyle\lrgrdel$}\hbox to\wd0
                        {\hfil$\scriptscriptstyle\rangle  $\hfil}}}}%
\def\parldel {\mathchoice
   {\mathopen {\setbox0=\hbox{$\displaystyle     \lrgldel$}\hbox to\wd0
                        {\hfil$\displaystyle     [       $\hfil}}}%
   {\mathopen {\setbox0=\hbox{$\textstyle        \lrgldel$}\hbox to\wd0
                        {\hfil$\textstyle        [        $\hfil}}}%
   {\mathopen {\setbox0=\hbox{$\scriptstyle      \lrgldel$}\hbox to\wd0
                        {\hfil$\scriptstyle      [        $\hfil}}}%
   {\mathopen {\setbox0=\hbox{$\scriptscriptstyle\lrgldel$}\hbox to\wd0
                        {\hfil$\scriptscriptstyle[        $\hfil}}}}%
\def\parrdel {\mathchoice
   {\mathclose{\setbox0=\hbox{$\displaystyle     \lrgrdel$}\hbox to\wd0
                        {\hfil$\displaystyle     ]       $\hfil}}}%
   {\mathclose{\setbox0=\hbox{$\textstyle        \lrgrdel$}\hbox to\wd0
                        {\hfil$\textstyle        ]        $\hfil}}}%
   {\mathclose{\setbox0=\hbox{$\scriptstyle      \lrgrdel$}\hbox to\wd0
                        {\hfil$\scriptstyle      ]        $\hfil}}}%
   {\mathclose{\setbox0=\hbox{$\scriptscriptstyle\lrgrdel$}\hbox to\wd0
                        {\hfil$\scriptscriptstyle]        $\hfil}}}}%

\def\aprs #1{\aprldel #1\aprrdel}%
\def\cons #1{\conldel #1\conrdel}%

\def\consopt #1{#1}%
\def\pars #1{\parldel #1\parrdel}%
\def\seqs #1{\seqldel #1\seqrdel}%
\def\sqn  #1{{\turnstile #1}}%

\def\quadcm {\rlap{\quad,}}%
\def\rdx #1{\ColorRGB{0.0 0.4 0.0}{#1}}%

\def\depth#1{\mathop{\rm depth} #1}

\def\CN{{\bf\rm C}_{{\bf\rm N}^*}}
\def\vars{\mathop{\rm vars}}

\def\SM#1#2#3#4#5{\pars{\seqs{\pars{#1,#2,#4};#3},
                             \seqs{\neg #1;\pars{\neg #2,\neg #3,#5}}}}

\def\ec{\cons{\enspace}}

\def\drv#1#2#3#4#5{\xy
 \xygraph{[]!{0;<2pc,0pc>:}
 {#2}-@2^<>(.5){\strut#4}
      _<>(.5){\strut#3}[#5]
 {#1}                         }
\endxy}

\def\ProofDia#1#2#3#4#5{
\xy
\xygraph{[]!{0;<#5pc,0pc>:}
{ }*=<0pt>{}:@2{|-}^<>(.5){#3}
                                _<>(.5){#2}  [#4]
{#1             }                                 }
\endxy
}

\def\DiaForbiddenConf{
\begin{matrix}
  \vcenter{
  \xymatrix@R=14pt@C=14pt{
     b \ar@[parcolor]@{-}[d] \ar@[parcolor]@{-}[r] \ar@[aparcolor]@{~}[dr] & 
     \SmallSize{\neg b} \ar@[aparcolor]@{~}[dl] \ar@[parcolor]@{-}[d] \\
     a \ar@[parcolor]@{-}[r] & \SmallSize{\neg a}
  }} &
  \quad , \quad &
  \vcenter{
  \xymatrix@R=14pt@C=14pt{
     b \ar@[parcolor]@{-}[d] \ar@[parcolor]@{-}[r] \ar@[aparcolor]@{~}[dr] & 
     \SmallSize{\neg b} \ar@[seqcolor]@{<~}[dl] \ar@[parcolor]@{-}[d] \\
     a \ar@[parcolor]@{-}[r] & \SmallSize{\neg a}
  }} &
  \quad \hbox{\rm and} \quad &
  \vcenter{
  \xymatrix@R=14pt@C=14pt{
     b \ar@[parcolor]@{-}[d] \ar@[parcolor]@{-}[r] \ar@[seqcolor]@{~>}[dr] & 
     \SmallSize{\neg b} \ar@[seqcolor]@{<~}[dl] \ar@[parcolor]@{-}[d] \\
     a \ar@[parcolor]@{-}[r] & \SmallSize{\neg a}
  }} \quad .
\end{matrix}
}

\def\DiaHexagonConf{
\xymatrix@R=14pt@C=14pt{
   & a \ar@[parcolor]@{-}[dl]  \ar@[seqcolor]@{~>}[dd] \ar@[parcolor]@{-}[ddr]
       \ar@[parcolor]@{-}[drr] \ar@[parcolor]@{-}[r]
   & \SmallSize{\neg a}
       \ar@[parcolor]@{-}[dll] \ar@[parcolor]@{-}[ddl] \ar@[seqcolor]@{~>}[dd]
       \ar@[seqcolor]@{~>}[dr]
   & \\
b  \ar@[seqcolor]@{~>}[dr] \ar@[parcolor]@{-}[drr] \ar@[parcolor]@{-}[rrr]  
   &  &
   & \SmallSize{\neg b}
     \ar@[parcolor]@{-}[dll]  \ar@[parcolor]@{-}[dl]
   \\
   & c 
     \ar@[parcolor]@{-}[r]
   & \SmallSize{\neg c} &   
}
}

\def\DiaSquareOneA#1#2#3#4{
\xygraph{[]!{0;<2.3pc,0pc>:}
{{#1}\strut}*[white]\cir<.5pc>{} (-     @{~>}@[seqcolor][r]
{{#2}\strut}*[white]\cir<.5pc>{}( -     @[parcolor][u]
{{#3}\strut}*[white]\cir<.6pc>{}  -   @[parcolor][l]
{{#4}\strut}*[white]\cir<.5pc>{}, -@[parcolor][ul]
{ \strut}*[white]\cir<.5pc>{}),-@{~>}@[seqcolor][ur]
{ \strut}*[white]\cir<.6pc>{} ,-     @[parcolor][u]
{ \strut}*[white]\cir<.5pc>{} )                         }
}

\def\DiaSquareOneB#1#2#3#4{
\xygraph{[]!{0;<2.3pc,0pc>:}
{{#1}\strut}*[white]\cir<.5pc>{} (-     @{~>}@[seqcolor][r]
{{#2}\strut}*[white]\cir<.5pc>{}( -     @[parcolor][u]
{{#3}\strut}*[white]\cir<.6pc>{}  -   @[parcolor][l]
{{#4}\strut}*[white]\cir<.5pc>{}, -@{--}@[antiparcolor][ul]
{ \strut}*[white]\cir<.5pc>{}),-@{~>}@[seqcolor][ur]
{ \strut}*[white]\cir<.6pc>{} ,-     @[parcolor][u]
{ \strut}*[white]\cir<.5pc>{} )                         }
}

\def\DiaSquareTwoA#1#2#3#4{
\xygraph{[]!{0;<2.3pc,0pc>:}
{{#1}\strut}*[white]\cir<.5pc>{} (-     @[parcolor][r]
{{#2}\strut}*[white]\cir<.5pc>{}( -     @[parcolor][u]
{{#3}\strut}*[white]\cir<.5pc>{}  -   @{~>}@[seqcolor][l]
{{#4}\strut}*[white]\cir<.5pc>{}, -@{~>}@[seqcolor][ul]
{ \strut}*[white]\cir<.5pc>{}),-@[parcolor][ur]
{ \strut}*[white]\cir<.5pc>{} ,-     @[parcolor][u]
{ \strut}*[white]\cir<.5pc>{} )                         }
}

\def\DiaSquareTwoB#1#2#3#4{
\xygraph{[]!{0;<2.3pc,0pc>:}
{{#1}\strut}*[white]\cir<.5pc>{} (-     @[parcolor][r]
{{#2}\strut}*[white]\cir<.5pc>{}( -     @[parcolor][u]
{{#3}\strut}*[white]\cir<.5pc>{}  -   @{~>}@[seqcolor][l]
{{#4}\strut}*[white]\cir<.5pc>{}, -@{~>}@[seqcolor][ul]
{ \strut}*[white]\cir<.5pc>{}),-@{--}@[antiparcolor][ur]
{ \strut}*[white]\cir<.5pc>{} ,-     @[parcolor][u]
{ \strut}*[white]\cir<.5pc>{} )                         }
}

\def\DiaSquareThreeA#1#2#3#4{
\xygraph{[]!{0;<2.3pc,0pc>:}
{{#1}\strut}*[white]\cir<.5pc>{} (-     @{~>}@[seqcolor][r]
{{#2}\strut}*[white]\cir<.5pc>{}( -     @[parcolor][u]
{{#3}\strut}*[white]\cir<.5pc>{}  -   @{~>}@[seqcolor][l]
{{#4}\strut}*[white]\cir<.5pc>{}, -@[parcolor][ul]
{ \strut}*[white]\cir<.5pc>{}),-@[parcolor][ur]
{ \strut}*[white]\cir<.5pc>{} ,-     @[parcolor][u]
{ \strut}*[white]\cir<.5pc>{} )                         }
}

\def\DiaSquareThreeB#1#2#3#4{
\xygraph{[]!{0;<2.3pc,0pc>:}
{{#1}\strut}*[white]\cir<.5pc>{} (-     @{~>}@[seqcolor][r]
{{#2}\strut}*[white]\cir<.5pc>{}( -     @{--}@[antiparcolor][u]
{{#3}\strut}*[white]\cir<.5pc>{}  -   @{~>}@[seqcolor][l]
{{#4}\strut}*[white]\cir<.5pc>{}, -@[parcolor][ul]
{ \strut}*[white]\cir<.5pc>{}),-@[parcolor][ur]
{ \strut}*[white]\cir<.5pc>{} ,-     @[parcolor][u]
{ \strut}*[white]\cir<.5pc>{} )                         }
}

\def\DiaSqProSeqOne{\xy
\xygraph{[]!{0;<2.3pc,0pc>:}
{a\strut}*      [white]\cir<.5pc>{}(-@{~>}@[seqcolor][u]
{b\strut}*      [white]\cir<.5pc>{}                  [d]
{ \strut}*=<1pc>[white]\frm{}       -@{~>}@[seqcolor][ul]
{d\strut}*=<1pc>[white]\frm{}      ,                 [l]
{c\strut}*      [white]\cir<.5pc>{} -@{~>}@[seqcolor][u] 
{ \strut}*      [white]\cir<.5pc>{})                     }
\endxy
}

\def\DiaSqProSeqTwo{\xy
\xygraph{[]!{0;<1.15pc,0pc>:}
{d\strut}*        [white]\cir<.5pc>{} -@{<~}@[seqcolor][dd]
{c\strut}*        [white]\cir<.5pc>{}                  [ddr]
{a\strut}*=<1.2pc>[white]\frm{}      (-@{~>}@[seqcolor][uul]
{ \strut}*=<1.2pc>[white]\frm{}      ,-@{~>}@[seqcolor][uur]
{b\strut}*=<1.2pc>[white]\frm{}      )                       }
\endxy
}

\def\DiaSqProSeqThree{\xy
\xygraph{[]!{0;<2.3pc,0pc>:}
{a\strut}*[white]\cir<.5pc>{}-@{~>}@[seqcolor][u]
{b\strut}*[white]\cir<.5pc>{}-@{~>}@[seqcolor][u]
{c\strut}*[white]\cir<.5pc>{}-@{~>}@[seqcolor][u]
{d\strut}*[white]\cir<.5pc>{}                    }
\endxy
}

\def\DiaSqProSeqFour{\xy
\xygraph{[]!{0;<1.15pc,0pc>:}
{a\strut}*        [white]\cir<.5pc>{} -@{~>}@[seqcolor][uu]
{b\strut}*        [white]\cir<.5pc>{}                  [uul]
{d\strut}*=<1.2pc>[white]\frm{}      (-@{<~}@[seqcolor][ddr]
{ \strut}*=<1.2pc>[white]\frm{}      ,-@{<~}@[seqcolor][ddl]
{c\strut}*=<1.2pc>[white]\frm{}      )                       }
\endxy
}

\def\DiaSqProSeqFive{\xy
\xygraph{[]!{0;<1.15pc,0pc>:}
{a\strut}*=<1.2pc>[white]\frm{}      (-@{~>}@[seqcolor][ruu]
{b\strut}*=<1.2pc>[white]\frm{}      ,-@{~>}@[seqcolor][luu]
{d\strut}*=<1.2pc>[white]\frm{}      ,                 [dd]
{c\strut}*        [white]\cir<.5pc>{} -@{~>}@[seqcolor][uu]
{ \strut}*        [white]\cir<.5pc>{})                      }
\endxy
}

\def\DiaSqProSeqSix{\xy
\xygraph{[]!{0;<2.3pc,0pc>:}
{a\strut}*      [white]\cir<.5pc>{}(-@{~>}@[seqcolor][u]
{b\strut}*      [white]\cir<.5pc>{}                  [d]
{ \strut}*=<1pc>[white]\frm{}       -@{~>}@[seqcolor][ul]
{d\strut}*=<1pc>[white]\frm{}      ,                 [l]
{c\strut}*      [white]\cir<.5pc>{} -@{~>}@[seqcolor][u] 
{ \strut}*      [white]\cir<.5pc>{}                  [d]
{ \strut}*=<1pc>[white]\frm{}       -@{~>}@[seqcolor][ur]
{ \strut}*=<1pc>[white]\frm{}      )                     }
\endxy
}

\def\DiaSqProSeqSeven{\xy
\xygraph{[]!{0;<1.15pc,0pc>:}
{d\strut}*=<1.2pc>[white]\frm{}      (-@{<~}@[seqcolor][ldd]
{c\strut}*=<1.2pc>[white]\frm{}      ,-@{<~}@[seqcolor][rdd]
{a\strut}*=<1.2pc>[white]\frm{}      ,                 [uu]
{b\strut}*        [white]\cir<.5pc>{} -@{<~}@[seqcolor][dd]
{ \strut}*        [white]\cir<.5pc>{})                      }
\endxy
}

\def\DiaSqProParOne{
\xy
\xygraph{[]!{0;<2.3pc,0pc>:}
{a\strut}*[white]\cir<.5pc>{}(-@[parcolor][r]
{b\strut}*[white]\cir<.5pc>{},-@[parcolor][u]
{d\strut}*[white]\cir<.5pc>{} -@[parcolor][r]
{c\strut}*[white]\cir<.5pc>{})               }
\endxy
}

\def\DiaSqProParTwo{
\xy
\xygraph{[]!{0;<2.3pc,0pc>:}
{a\strut}*[white]\cir<.5pc>{}(-@[parcolor][r]
{b\strut}*[white]\cir<.5pc>{},-@[parcolor][u]
{d\strut}*[white]\cir<.5pc>{} -@[parcolor][r]
{c\strut}*[white]\cir<.5pc>{},-@[parcolor][ur]
{ \strut}*[white]\cir<.5pc>{})                }
\endxy
}

\def\DiaSqProParThree{
\xy
\xygraph{[]!{0;<2.3pc,0pc>:}
{a\strut}*[white]\cir<.5pc>{}(-@[parcolor][r]
{b\strut}*[white]\cir<.5pc>{} -@[parcolor][u]
{ \strut}*[white]\cir<.5pc>{},-@[parcolor][u]
{d\strut}*[white]\cir<.5pc>{} -@[parcolor][r]
{c\strut}*[white]\cir<.5pc>{})                }
\endxy
}

\def\DiaSqProParFour{
\xy
\xygraph{[]!{0;<2.3pc,0pc>:}
{a\strut}*[white]\cir<.5pc>{}(-@[parcolor][r]
{b\strut}*[white]\cir<.5pc>{} -@[parcolor][ul]
{ \strut}*[white]\cir<.5pc>{},-@[parcolor][u]
{d\strut}*[white]\cir<.5pc>{} -@[parcolor][r]
{c\strut}*[white]\cir<.5pc>{})                }
\endxy
}

\def\rdx #1{\underline{#1\phantom{\hbox to0pt{,\hss}}}}

\theoremstyle{plain}

\def\doi{2 (2:4) 2006}
\lmcsheading%
{\doi}
{24}
{}
{}
{Jul.~\phantom{0}5, 2005}
{Apr.~\phantom{0}3, 2006}
{}
\begin{document}

\title[]{A System of Interaction and Structure II:\\ The Need for Deep
Inference} 

\author[A.~Tiu]{Alwen Tiu} 
\address{Computer Science Laboratory,
Research School in Information Science and Engineering,
Australian National University, Canberra ACT 0200, Australia.} 
\email{Alwen.Tiu@rsise.anu.edu.au}
\thanks{A major part of this work was done when the author was at
Fakult\"at Informatik, TU Dresden. The author received support from
INRIA Lorraine (LORIA) during the completion of this work.}

\keywords{proof theory, deep inference, sequent calculus, calculus of
structures, non-commutative logics}
\subjclass{F.4.1}


\begin{abstract}
  This paper studies properties of the logic $\BV$, which is an
  extension of multiplicative linear logic (MLL) with a self-dual
  non-commutative operator. $\BV$ is presented in the {\em calculus of
  structures}, a proof theoretic formalism that supports {\em deep
  inference}, in which inference rules can be applied anywhere inside
  logical expressions.  The use of deep inference results in a simple
  logical system for MLL extended with the self-dual non-commutative
  operator, which has been to date not known to be expressible in
  sequent calculus. In this paper, deep inference is shown to be
  crucial for the logic $\BV$, that is, any restriction on the
  ``depth'' of the inference rules of $\BV$ would result in a strictly
  less expressive logical system.
\end{abstract}

\maketitle
\vskip-\bigskipamount

\section{Introduction}

This paper is the second part of a planned series of papers on {\em the calculus
of structures}, started in \cite{GugV02}. 
The calculus of structures is a new formalism for presenting logics. 
It is a generalization of the one-sided sequent calculus 
(also known as Gentzen-Sh\"utte sequent calculus \cite{TrolBasic96}). 
One of the main features of the calculus of structures is that it allows
{\em deep inference}, i.e., inference rules in the calculus of structures
can be applied arbitrarily deep inside logical expressions.
This is in contrast with the traditional logical rules in sequent calculus,
as in, e.g., Gentzen's LK \cite{GenILD35} or Girard's linear logic \cite{GirTCS87}, 
where the logical rules manipulate only the topmost connectives of the formulas.
In \cite{GugV02}, and several other related works (e.g., \cite{BruTiu2001,GugStr01,Bru03phd,StraThesis}), 
it is shown that deep inference gives rise to a number of interesting
proof theoretical properties for the logical systems formalized in the 
calculus of structures that are not observable in systems without deep inference. 
However, if one is concerned only with provability of logical expressions,
deep inference might seem to introduce unnecessary non-determinism into 
{\em proof search} (although this problem has been partly addressed 
in \cite{GugV02}, using a technique called {\em splitting}). 
The work presented in this paper had started with
the conjecture that deep inference is not necessary for provability; 
a conjecture which turns out to be wrong. We present a counterexample, 
a class of logical expressions in System $\BV$~\cite{GugV02}, an extension of 
multiplicative linear logic with a self-dual non-commutative connective, 
which is provable only if there is no fixed bound to the depth of the applicability
of the rules in $\BV$, and hence rule out any candidate ``shallow system'' for
$\BV$. 

The main result of this paper can be seen as a first technical argument to the claim 
that deep inference is a non-trivial extension to the traditional sequent calculus,
in the sense that it allows for a simple formulation of a logic, 
which does not admit any straightforward formulation in sequent calculus 
without deep inference. By ``sequent calculus without deep inference'' we have in mind sequent
systems with list-like contexts and whose logical rules manipulate only the topmost connectives. 
We regard such systems as a subset of {\em shallow systems}, the precise definition
of which is given in Section~\ref{sec:shallow}.
There are calculi other than the calculus of structures which also employ some sort of deep inference, 
notably, the logic of bunched implication~\cite{Ohearn99bsl} 
and the display calculus~\cite{BelnapJPL82},  where the context of the sequent
is a tree-like structure and inference rules can access any part of the tree.
But there has been so far no technical analyses in those calculi concerning 
the necessity of deep inference.

We hope also to shed light on the known open problem of the 
sequentialization of Retor\'e's Pomset logic \cite{RetTLCA97}.
Pomset logic is an extension of multiplicative linear logic with a
self-dual non-commutative connective. It is presented in the calculus of
proofnets~\cite{GirTCS87}, and there has to date been no known sequentialization theorem yet.
The logic $\BV$ and Pomset logic are closely related; in fact, there is
a sound translation from $\BV$ to Pomset logic preserving provability \cite{StraThesis}.
We conjecture that they are actually the same logic.
The result on the necessity of deep-inference of $\BV$ therefore explains 
to some extent the difficulty in the sequentialization of Pomset logic. 

This paper is organised as follows. In Section~\ref{sec:structures}, we define {\em structures}, the analog
to sequents in sequent calculus. This is then followed by a presentation of System $\BV$ 
in Section~\ref{sec:bv}. Section~\ref{sec:overview} gives an overview of the main idea behind the proof of the 
necessity of deep inference, that is, the construction of a class of provable structures
in $\BV$ which serve as the counterexample that rules out any candidate shallow system for $\BV$.   
The formal proof for the necessity of deep inference relies on a graph representation of
structures, called the {\em relation webs}~\cite{GugV02}, 
which is reviewed in Section~\ref{sec:relweb}.
Section~\ref{sec:shallow} gives a technical definition of shallow systems. 
In Section~\ref{sec:deep} we prove formally the necessity of deep-inference
for System $\BV$: for every candidate shallow system for $\BV$,
there is a structure provable in $\BV$ but not provable in that candidate system. 
Section~\ref{sec:conc} concludes the paper.

\section{Structures}
\label{sec:structures}

The logical expressions in the calculus of structures are called {\em structures}. 
A structure can be seen as representing a class of logical formulas, 
where each logical connective corresponds to a unique structure-forming
relation, or {\em structural relation}. 
This is similar to the reading of sequents in sequent calculus:
a (one-sided) sequent $\vdash A, B$ in linear logic can be seen as representing
both the formulas $A \lpar B$ and $B \lpar A$, interpreting the 
`,' (comma) in the sequent as $\lpar$. 
However, unlike sequent calculus, since each connective is assigned 
its own structural relation, the distinction between
connectives and strutural relations are not particularly
relevant if we consider only the combinatorial properties of
the logical systems in the calculus of structures (e.g., permutability,
cut-elimination, etc.). 
We are therefore contend with having just structures as the only syntactical
expressions of the logical systems in calculus of structures.
The class of structures we are interested in is the following.

\begin{defi}
\label{DefStructures}
There are infinitely many {\em positive atoms}
and {\em negative atoms}.  Atoms, positive or negative, are denoted 
by $a$, $b$, \dots.
{\em Structures} are denoted by $S$, $P$, $Q$, $R$, $T$, $U$ and $V$.  The
structures of the language $\BV$ are generated by
$$
S \grammareq
             a                                        \mid
             \un                                      \mid
             \pars{\,\underbrace{S,\dots,S}_{{}>0}\,} \mid
             \aprs{\,\underbrace{S,\dots,S}_{{}>0}\,} \mid
             \seqs{\,\underbrace{S;\dots;S}_{{}>0}\,} \mid
             \neg S                                       \quadcm
$$
where $\un$, the {\em unit}, is not an atom; ${\pars{S_1,\dots,S_h}}$ is a 
{\em par structure}, ${\aprs{S_1,\dots,S_h}}$ is a {\em copar structure} and
${\seqs{S_1;\dots;S_h}}$ is a {\em seq structure}; ${\bar S}$ is the {\em negation} of
the structure $S$.  Structures with a hole not occuring in the scope of a negation
are denoted by ${S\cons{\enspace}}$. 
The structure $R$ is a {\em substructure} of $S\cons R$, and $S\cons\enspace$ is its
{\em context}.  
\end{defi}
The par and co-par structural relations correspond to the $\lpar$ and the $\ltens$
connectives in linear logic~\cite{GirTCS87}, while the seq-relation is a self-dual 
non-commutative connective. 
We often omit the curly braces in a structure context in
cases where structural parentheses fill the hole exactly: 
for example, we shall write $S\pars{R,T}$ to mean $S\cons{\pars{R,T}}$. 

\begin{defi}
The structures of the language $\BV$ are equivalent modulo the
relation $=$, defined in Figure \ref{figSE}.  There, $\vec R$,
$\vec T$ and $\vec U$ stand for finite, non-empty sequences of structures
(sequences may contain `$,$' or `$;$' separators as appropriate in the context).
A structure $S$ is in {\em normal form} when either $S = \un$ or 
there is no unit $\un$ appearing in it and the only negated structures appearing 
in it are atoms. 
Similarly, a structure context $S\ec$ is in normal form when 
there is no unit $\un$ appearing in it and the only negated structures 
appearing in it are atoms. 
Given a structure $S$, we talk about the atom occurrences of $S$ when we consider
all the atoms appearing in $S$ as distinct. Therefore, in the structure $\seqs{a;a}$ 
there are two different occurrences of the atom $a$. The set of all the atom 
occurrences in $S$ is denoted with $\occ S$. 
\end{defi}

\begin{figure}
{
$$\vcenter{\hbox{\vtop{\halign{\hfil#\hfil\cr
{\bf Associativity\strut}\cr
\noalign{\medskip}%
\hbox{$\begin{array}{rl}
\smash{\pars{\vec R,\pars{\vec T}}}        &= 
                              \smash{\pars{\vec R,\vec T}}       \\
\smash{\aprs{\vec R,\aprs{\vec T}}}        &= 
                              \smash{\aprs{\vec R,\vec T}}       \\
\smash{\seqs{\vec R;\seqs{\vec T};\vec U}} &= 
                              \smash{\seqs{\vec R;\vec T;\vec U}}\\
\end{array}$}\cr
\noalign{\bigskip}%
{\bf Unit\strut}\cr
\noalign{\medskip}%
\hbox{$\begin{array}{rl}
\smash{\pars{\un,\vec R}}                             &= 
                                            \smash{\pars{\vec R}}\\
\smash{\aprs{\un,\vec R}}                             &= 
                                            \smash{\aprs{\vec R}}\\
\smash{\seqs{\un;\vec R}} = \smash{\seqs{\vec R;\un}} &= 
                                            \smash{\seqs{\vec R}}\\
\end{array}                                            
$}\cr
\noalign{\bigskip}%
{\bf Singleton\strut}\cr
\noalign{\medskip}%
\hbox{$\smash{\pars{R}} = \smash{\aprs{R}} = \smash{\seqs{R}} = R$}\cr
}}
$\qquad \qquad$
\enspace\vtop{\halign{\hfil#\hfil\hfil\cr
{\bf Commutativity\strut}\cr
\noalign{\medskip}%
\hbox{$\begin{array}{rl}
\smash{\pars{\vec R,\vec T}} &= \smash{\pars{\vec T,\vec R}}\\
\smash{\aprs{\vec R,\vec T}} &= \smash{\aprs{\vec T,\vec R}}\\
\end{array}$}\cr
\noalign{\bigskip}%
{\bf Negation\strut}\cr
\noalign{\medskip}%
\hbox{$\begin{array}{rl}
\smash{\neg\un}                               &= 
                           \smash{\un}                           \cr
\smash{\overline{\strut\pars{R_1,\dots,R_h}}} &= 
                           \smash{\aprs{\neg R_1,\dots,\neg R_h}}\cr
\smash{\overline{\strut\aprs{R_1,\dots,R_h}}} &= 
                           \smash{\pars{\neg R_1,\dots,\neg R_h}}\cr
\smash{\overline{\strut\seqs{R_1;\dots;R_h}}} &= 
                           \smash{\seqs{\neg R_1;\dots;\neg R_h}}\cr
\smash{\skew3\neg{\neg R}}                    &= 
                           \smash{R}                             \cr
\end{array}$}\cr
\noalign{\bigskip}%
{\bf Contextual Closure\strut}\cr
\noalign{\medskip}%
\hbox{if $R=T$ then $S\cons R=S\cons T$}\cr
}}}}
$$\quad\hbox{$\vcenter{\hbox{%
}}$}
}
\caption{Syntactic equivalence\/ $=$}
\label{figSE}
\end{figure}

We shall now define precisely the depth of a structure context.
This definition will come handy later when proving various properties related
to the depth of inference rules.
Note that since structures and structure contexts are considered modulo associativity 
and commutativity (for par and co-par), 
the depth of a structure context is also considered modulo these equalities.
In its normal form, a structure can be viewed as a finitely branching tree. 
A structure context is then a particular tree with a hole $\ec$ as a leaf. 
The depth is measured as the length of the branch ending with $\ec$. 
The formal definition is as follows.

\begin{defi}
\label{DepthOfContext}
Let $S\ec$ be a normal structure context. 
The {\em depth} of $S\ec$ is defined as follows:
\begin{enumerate}
\item $\depth~{\ec} = 0$, 
\item $\depth~{\pars{S_1, S_2\ec}} = 
         \begin{cases}
           \depth\,{S_2\ec}, & \mbox{if $S_2\ec=\pars{S_2', S_2''\ec}$},\cr
           \depth\,{S_2\ec}+1, & \mbox{otherwise},
         \end{cases}$
\item $\depth~{\aprs{S_1, S_2\ec}} = 
         \begin{cases}
           \depth\,{S_2\ec}, & \mbox{if $S_2\ec=\aprs{S_2', S_2''\ec}$},\cr
           \depth\,{S_2\ec}+1, & \mbox{otherwise},
         \end{cases}$
\item $\depth~{\seqs{S_1; S_2\ec}} =
         \begin{cases}
            \depth\,{S_2\ec}, & \mbox{if $S_2\ec=\seqs{S_2';S_2''\ec}$} \cr
                            & \mbox{or $S_2\ec=\seqs{S_2'\ec;S_2''}$},\cr
            \depth\,{S_2\ec}+1, & \mbox{otherwise},
         \end{cases}$
\item $\depth~{\seqs{S_1\ec; S_2}} =
         \begin{cases}
            \depth\,{S_1\ec}, & \mbox{if $S_1\ec=\seqs{S_1';S_1''\ec}$} \cr
                            & \mbox{or $S_1\ec=\seqs{S_1'\ec;S_1''}$},\cr
            \depth\,{S_1\ec}+1, & \mbox{otherwise.}
         \end{cases}$
\end{enumerate}
\end{defi}

For example, the structure context $\pars{a, b, \ec}$ has depth 1 and
$\pars{\seqs{\ec;c}, \seqs{b;c}}$ has depth 2. 
A structure $R$ is said to {\em occur at depth $n$} in $S\cons R$ if
$n$ is the depth of $S\ec$.

\section{System $\BV$}
\label{sec:bv}

\begin{figure}
$$
\begin{array}{ccccc}
\inf{\atir}
    {S\pars{a,\neg a}}
    {S\cons \un}                      &
\inf{\aatir}
    {S\cons \un}
    {S\aprs{a,\neg a}}          &      
\inf{\seqr}
    {S\pars{\seqs{R;R'},\seqs{T;T' }}}
    {S\seqs{\pars{R,T} ;\pars{R',T'}}} &
\inf{\aseqr}
    {S\seqs{\aprs{R,R'};\aprs{T ,T'}}}
    {S\aprs{\seqs{R;T },\seqs{R';T'}}}  &     
\inf{\swir}
    {S\pars{\aprs{R,R'},T }}
    {S\aprs{\pars{R,T },R'}}           
\end{array}    
$$
\caption{System\/ $\SBV$}
\label{figBV}
\end{figure}

In this section, we review some key concepts and terminologies in the calculus of structures 
and present System $\BV$. 

Recall that we associate to each connective a structural relation. This has a consequence
on the shape of inference rules in the calculus of structures: 
there is basically no need for branching inference rules, i.e.,
inference rules with more than one premise. 
Derivations in the calculus of structures are sequences of inference rules,
instead of tree-shape structures. 
Consider for example the following {\em tensor} rule of linear logic
in sequent calculus:
$$
\vcenter{\iinf{\ltens}
              {\sqn{A\ltens B,\Phi,\Psi}}
              {\sqn{A        ,\Phi     }}
              {\sqn{        B,     \Psi}}}
\quad.
$$
In the calculus of structure, we would present the rule as something 
like
$$
\inf{\ltens}{\pars{\aprs{A, B}, \Phi, \Psi}}{\aprs{\pars{A,\Phi},\pars{B,\Psi}}}
$$
Here we use the par structural relation to interpret
$\lpar$ and the co-par to interpret $\ltens$. 
This simulation of the tensor rule in the calculus of structures
can be generalized to arbitrary one-sided sequent systems. This idea has been
used to simulate several known sequent systems inside the calculus of structures
(see e.g., \cite{BruTiu2001,StraThesis}), through which the cut-elimination results 
are obtained for the corresponding systems in  the calculus of structures.

Another important feature of the calculus of structures is that it allows {\em deep inference}.
Inference rules can be applied to any substructure of a given structure. Thus,
inference rules in the calculus of structures can be seen as rewrite rules on structures.
This is in contrast to traditional sequent systems where introduction rules are usually 
applied to the topmost connectives of formulas.
Typical rules in the calculus of structures have the form:
$$
\inf{\rho}{S\cons{R}}{S\cons{T}}
$$
where $\rho$ is the name of the rule, $R$ and $T$ are structures and $S\ec$ is 
a structure context.
Common to all systems already studied in the calculus of structures are the interaction rules:
$$
\vcenter{\inf{\intr}{S\pars{R,\overline{R}}}{S\cons{\un}}}
\quad \hbox{\rm and} \quad
\vcenter{\inf{\aintr}{S\cons{\un}}{S\aprs{R,\overline{R}}}}
$$
which correspond to the identity rule and the cut rule in sequent calculus, respectively.

We shall now define precisely System $\BV$ with its notions of inference rules, derivations
and proofs.

\begin{defi} 
An {\em inference rule} is a scheme
$
\vcenter{\infnote{\rho}
                 {R}
                 {T}{}}
$,
where $\rho$ is the {\em name} of the rule, $T$ is its {\em premise} and $R$ is its
{\em conclusion}.  Rule names are denoted by $\rho$ and $\pi$.  A
{\em formal system}, denoted by $\sysS$, is a set of rules.  A
{\em derivation} in a system $\sysS$ is a finite chain of instances of
rules of $\sysS$, and is denoted by $\Delta$.  A derivation can consist of just
one structure.  The topmost structure in a derivation is called its
{\em premise}; the lowest structure is called {\em conclusion}.  A derivation 
$\Delta$ whose premise and conclusion are $T$ and $R$, and whose rules are
in $\sysS$ is denoted by
$\vcenter{
\xy
\xygraph{[]!{0;<2pc,0pc>:}
{T}-@2^<>(.5){\strut\sysS}
      _<>(.5){\strut\Delta}[d]
{R}                         }
\endxy} .
$
An instance of a rule $\vcenter{\inf{\rho}{R}{T}}$ is 
called {\em non-trivial} if $R \not = T$, otherwise it is {\em trivial}.
\end{defi}
     
\begin{defi}
System $\SBV$ \cite{GugStr01} is shown in Figure~\ref{figBV}. 
The rules in $\SBV$ are of the form $\vcenter{\inf{\rho}{S\cons{R}}{S\cons{T}}}$.
The structure $R$ is called {\em redex} and the structure $T$ is called {\em contractum}. 
\end{defi}

The rules $\atir$ and $\aatir$ are the atomic versions of the more general interaction rules 
$\intr$ and $\aintr$. The non-atomic interaction rules can be reduced to the atomic ones, 
as shown in \cite{GugStr01}.
System $\SBV$ consists of a down fragment $\{\atir,\swir,\seqr\}$ and 
an up fragment $\{\aatir,\swir,\aseqr\}$. 
The rules $\atir$ and $\aseqr$ in the up fragment are admissible (for proofs) as a consequence
of the cut elimination theorem for this system \cite{GugV02}. 
Therefore, it is enough to study the down fragment for its 
proof search properties. We need an extra rule, a logical axiom rule, to define
proofs in $\SBV$.

\begin{defi} 
The down fragment of $\SBV$, together with the rule
$\smash{
\vcenter{\infnote{\unr}
                 {\enspace\un\enspace}
                 {}{}}
}$,   
is called System $\BV$ (Figure~\ref{figBVP}). 
\end{defi}

\begin{figure}
$$
\inf{\unr}
    {\enspace\un\enspace}
    {}
\qquad
\inf{\atir}
    {S\consopt{\pars{a,\neg a}}}
    {S\cons   {\un}}
\qquad
\inf{\swir}
    {S\consopt{\pars{\aprs{R,R'},T}}}
    {S\consopt{\aprs{\pars{R,T},R'}}}
\qquad
\inf{\seqr}
    {S\consopt{\pars{\seqs{R;R'},\seqs{T;T' }}}}
    {S\consopt{\seqs{\pars{R,T} ;\pars{R',T'}}}}
$$    
\caption{System $\BV$}
\label{figBVP}
\end{figure}

\begin{defi} A {\em proof}, denoted by $\Pi$, is a derivation 
whose top is an instance of $\unr$. A system $\sysS$ {\em proves} $R$ if 
there is in
$\sysS$ a proof $\Pi$ whose conclusion is $R$, written
$\vcenter{
\xy
\xygraph{[]!{0;<1.5pc,0pc>:}
{ }*=<0pt>{}:@2{|-}^<>(.5){\sysS}
                                _<>(.5){\Pi}  [d]
{R             }                                 }
\endxy
}.$ 
The {\em length} of a proof $\Pi$ is the number of occurrences of
inference rules in $\Pi$.
Two systems are {\em equivalent} if they prove the same structures.  
\end{defi}

Given a provable structure in $\BV$, we can remove all the dual instances of atoms, except 
maybe for some pairs, and get another provable structure. 
In this way, we can see whether there are certain local properties obeyed 
by all provable structures. This will be useful later when proving the necessity
of deep inference (see Section~\ref{sec:deep}).
\begin{prop}
\label{PropProofSubst}
If there is a derivation 
$\vcenter{
  \infnote
     {\atir}
     {\drv{R}{T\pars{a,\neg a}}{\Delta}{\BV}{d} }
     {T\cons{\un}}
     {}
}$
then there is a derivation 
$\vcenter{ \drv{R'}{T\cons{\un}}{\Delta'}{\BV}{d} }$
where $R'$ and $\Delta'$ are obtained from $R$ and $\Delta$, respectively, 
by replacing the atom occurrences $a$ and $\neg a$ with units.
\end{prop}
\proof
By simple induction on the length of $\Delta$.
\qed

\section{Overview of the Counterexample}
\label{sec:overview}

The counterexample is based on the structure
$$\pars{\seqs{\pars{a,b};c}, \seqs{\neg a;\pars{\neg b,\neg c}}},$$
which we shall refer to as $S_0$. 
In order to prove the structure, all dual pairs of atoms must be brought into
the same par substructures so that the interaction rule $\atir$ can be applied. It is quite obvious 
that more than one application of inference rules in $\BV$ is needed to 
bring the dual pairs of atoms into the same par substructures. 
For instance, to bring $a$ and $\neg a$ into the same substructure (via the 
$\seqr$ rule), the par substructure $\pars{a,b}$ needs to be changed to $\seqs{a;b}$ first. 
The same observation applies to the substructure $\pars{\neg b, \neg c}$: it must be changed to $\seqs{\neg b; \neg c}$
first before $c$ can be brought into the same par substructure with $\neg c$. 
It can be shown that $S_0$ has the following property: all of its proofs must begin with (reading the
proofs bottom up) either the substructure $\pars{a,b}$ or $\pars{\neg b, \neg c}$ as the redex.
A proof of $S_0$ is given in Figure~\ref{fig:S0}, where the first redex is the substructure
$\pars{a,b}$ (modulo structure equality). The underlined substructures in the proof
are redexes of the rules. The equality sign in the rules indicates applications of structure equalities. 
There is another proof of $S_0$ which is not shown and which starts with $\pars{\neg b, \neg c}$ as
the redex. The substructures $\pars{a, b}$ and $\pars{\neg b, \neg c}$ occur at depth
2 in $S_0$. Therefore this structure is provable in a given candidate shallow system for $\BV$ 
only if the proof system allows for rules which apply
beyond depth 2 in the structure (the notion of the depth of inference rules will be made
precise later). 

\begin{figure}
$$
\dernote{=}{\quad.                  }
        {\pars{
            \seqs{\rdx {\pars{a, b}} ; c},
            \seqs{\neg a ; \pars{\neg b, \neg c}}
          }          } {
\root   {\seqr}
        {\pars{
            \seqs{\rdx {\pars{\seqs{a ; \un}, \seqs{\un ; b}}} ; c}, 
            \seqs{\neg a ; \pars{\neg b, \neg c}}  
          } }{
\root   {=}
        {\pars{
            \rdx {\seqs{\seqs{\pars{a, \un} ; \pars{\un, b}};c}},
            \seqs{\neg a;\pars{\neg b,\neg c}}}
        } {
\root   {\seqr}{\rdx {\pars{
                  \seqs{a;b;c},
                  \seqs{\neg a;\pars{\neg b,\neg c}}}}
               }{
\root   {\atir}{\seqs{\rdx{\pars{a,\neg a}};
                \pars{\seqs{b;c},\neg b,\neg c}}}{
\root   {=}
        {\rdx {\seqs{\un ; \pars{\seqs{b;c},\neg b,\neg c}}}}{
\root   {=}{\pars{\seqs{b;c},\rdx {\neg b},\neg c}}{
\root   {\seqr}
        {\pars{\rdx {\seqs{b;c},\seqs{\neg b ; \un}}, \neg c} }{
\root   {=}
        {\rdx{\pars{\seqs{\pars{b,\neg b}; \pars{c, \un}},\neg c}}}{
\root   {\seqr}{\rdx{\pars{\seqs{\pars{b, \neg b};c},\seqs{\un;\neg c}}}}{
\root   {=}
        { \seqs{\rdx{\pars{b,\neg b, \un}};\pars{c,\neg c}} }{
\root   {\atir}{\seqs{\pars{b,\neg b};\rdx{\pars{c,\neg c}}}}{
\root   {\atir}{\seqs{\pars{\rdx{b, \neg b}} ; \un}}{
\root   {=}{\rdx{\seqs{\un ; \un}} }{
\root   {\unr}{\un}{
\leaf   {}}}}}}}}}}}}}}}}
$$
\caption{A proof of the $S_0$ structure}
\label{fig:S0}
\end{figure}

Another interesting property of the $S_0$ structure is that it can be superimposed with itself,
forming a bigger structure such that its proofs must begin with an innermost par substructure
as the redex. We illustrate here how such structures can be constructed, the formal definition
will be given in Section~\ref{sec:deep}. Consider the following ``parameterized' $S_0$:
$$
\pars{\seqs{\pars{a,b, R };c}, \seqs{\neg a;\pars{\neg b,\neg c, T }}}.
$$
We denote with $S_0\cons{U}\cons{V}$ the above structure with $U$ replacing $R$ and
$V$ replacing T. 
The following structure is a superimposition of three $S_0$ structures, which we call $S_1$:
$$
\pars{\seqs{S_0\cons{a_1}\cons{b_1};c_1}, 
      \seqs{\neg a_1;S_0\cons{\neg b_1}\cons{\neg c_1}}}.
$$
A proof of $S_1$ can be constructed from the schematic derivation of the parameterized
$S_0$ in Figure~\ref{fig:param-S0}. That is, we apply the derivation scheme 
to the substructures $S_0\cons{a_1}\cons{b_1}$ and $S_0\cons{\neg b_1}\cons{\neg c_1},$
yielding the structure
$\pars{\seqs{\pars{a_1,b_1};c_1}, \seqs{\neg a_1;\pars{\neg b_1,\neg c_1}}}.$
The latter is an $S_0$ structure (modulo renaming of atoms) and is therefore provable.  
Again as with $S_0$, it can be shown that any proof of $S_1$ must begin with
an innermost par substructure as the redex. In this case, the depth of
the redex is 4. The $S_1$ structure can be further superimposed with other $S_0$ structures
to produce bigger structures with the same property. A recursive procedure to produce 
such structures is given in Section~\ref{sec:deep}. As we shall see later, given
any $n$, this procedure generates a provable structure with the property that
its proofs must begin with a redex at depth $2n$. Thus given any candidate
shallow system for $\BV$, a provable structure can be generated such that its proofs
must start at a depth that is beyond the depth of the rules in that particular
shallow system, and hence cannot be proved in the system. 

\begin{figure}
$$
\dernote{=}{\quad.                  }
        {\pars{
            \seqs{\rdx {\pars{a, b, R}} ; c},
            \seqs{\neg a ; \pars{\neg b, \neg c, T}}
          }          } {
\root   {\seqr}
        {\pars{
            \seqs{\rdx {\pars{\seqs{a ; \un}, \seqs{\un ; \pars{b,R}}}} ; c}, 
            \seqs{\neg a ; \pars{\neg b, \neg c, T}}  
          } }{
\root   {=}
        {\pars{
            \rdx {\seqs{\seqs{\pars{a, \un} ; \pars{\un, b,R}};c}},
            \seqs{\neg a;\pars{\neg b,\neg c,T}}}
        } {
\root   {\seqr}{\rdx {\pars{
                  \seqs{a;\pars{b,R};c},
                  \seqs{\neg a;\pars{\neg b,\neg c,T}}}}
               }{
\root   {\atir}{\seqs{\rdx{\pars{a,\neg a}};
                \pars{\seqs{\pars{b,R};c},\neg b,\neg c,T}}}{
\root   {=}
        {\rdx {\seqs{\un ; \pars{\seqs{\pars{b,R};c},\neg b,\neg c,T}}}}{
\root   {=}{\pars{\seqs{\pars{b,R};c},\rdx {\pars{\neg b, T}},\neg c}}{
\root   {\seqr}
        {\pars{\rdx {\seqs{\pars{b,R};c},\seqs{\pars{\neg b,T} ; \un}}, \neg c} }{
\root   {=}
        {\rdx{\pars{\seqs{\pars{b,R, \neg b, T}; \pars{c, \un}},\neg c}}}{
\root   {\seqr}{\rdx{\pars{\seqs{\pars{b, \neg b,R,T};c},\seqs{\un;\neg c}}}}{
\root   {=}
        { \seqs{\rdx{\pars{b,\neg b,R,T, \un}};\pars{c,\neg c}} }{
\root   {\atir}{\seqs{\pars{b,\neg b,R,T};\rdx{\pars{c,\neg c}}}}{
\root   {\atir}{\seqs{\pars{\rdx{b, \neg b}, R,T} ; \un}}{
\root   {=}{\rdx{\seqs{\pars{\un, R, T} ; \un}} }{
\leaf   {\pars{R,T}}}}}}}}}}}}}}}}
$$
\caption{A derivation for the parameterized $S_0$}
\label{fig:param-S0}
\end{figure}

\section{Characterisation of Structures}
\label{sec:relweb}

In this section we consider a different representation for structures, 
a special kind of graph called {\em relation webs} \cite{GugV02} (originally 
called {\em traces} \cite{GugV99}), which is 
a generalisation of series-parallel orders \cite{Moeh89}. 
In this representation, structures can be characterized by the absence
of certain forbidden configurations in their relation webs. 
This characterisation is useful when we want to study general properties of 
inference rules. This section reviews the definitions and the main results 
concerning relation webs already presented in \cite{GugV99, GugV02}.

\def\seqrelS {{\seqrel_S}}%
\def\aseqrelS {{\aseqrel_S}}%
\def\parrelS {{\parrel_S}}%
\def\aparrelS {{\aparrel_S}}%
\def\wbS {\wb_S}%

Let $S$ be a structure in normal form. The {\em structural relations}
$\seqrelS$ ({\em seq}),
$\aseqrelS$ ({\em co-seq}), 
$\parrelS$ ({\em par}) and 
$\aparrelS$ ({\em co-par}) are defined as the smallest sets such that
${\seqrel_S},{\aseqrel_S},{\parrel_S},{\aparrel_S}\subset(\occ S)\times (\occ S)$ and, 
for every $S'\ec$, $U$ and $V$ and for every $a$ in $U$ and $b$ in $V$, the following
hold:
\begin{enumerate}
\item if $S=S'\seqs{U;V}$ then $a\seqrel_S b$ and $b\aseqrel_S a$;
\item if $S=S'\pars{U,V}$ then $a\parrel_S b$;
\item if $S=S'\aprs{U,V}$ then $a\aparrel_S b$.
\end{enumerate}
To a structure that is not in normal form we associate the structural relations
obtained from any of its normal forms.  The quadruple $(\occ
S,{\seqrel_S},{\parrel_S},{\aparrel_S})$ is called the relation web of $S$, written
$\wbS$. We shall omit the subscripts in ${\seqrel_S}$, ${\aseqrel_S}$, ${\parrel_S}$ and ${\aparrel_S}$ 
when it is clear from context which structure we refer to.
Given two sets of atom occurrences $\mu$ and $\nu$, we write
$\mu\seqrel\nu$, $\mu\aseqrel\nu$, $\mu\parrel\nu$ and $\mu\aparrel\nu$ to
indicate situations where, for every $a$ in $\mu$ and for every $b$ in $\nu$,
they hold, respectively, $a\seqrel b$, $a\aseqrel b$, $a\parrel b$ and
$a\aparrel b$.
For example, in $\aprs{\seqs{a,\neg b},\pars{\neg c,d}}$
we have the following relations: $a\seqrel\neg b$, $a\aparrel\neg c$, $a\aparrel
d$, $\neg b\aseqrel a$, $\neg b\aparrel\neg c$, $\neg b\aparrel d$, $\neg
c\aparrel a$, $\neg c\aparrel\neg b$, $\neg c\parrel d$, $d\aparrel a$,
$d\aparrel\neg b$, $d\parrel\neg c$.

The following proposition states that all the atoms in a sub-structure 
are in the same structural relations with respect to 
each of the atoms surrounding them:

\begin{prop} 
\label{prop:substructure}
\cite{GugV02}
Given a structure\/ $S\cons{R}$ and two atom occurrences\/ $a$ in\/
$S\ec$ and\/ $b$ in\/ $R$, if\/ $a\seqrel b$ (respectively, $a\aseqrel
b$, $a\parrel b$, $a\aparrel b$) then\/ $a\seqrel c$ (respectively, $a\aseqrel
c$, $a\parrel c$, $a\aparrel c$) for all the atom occurrences\/ $c$ in\/ $R$.
\end{prop}

\begin{thm}  
\label{TheoStructRel} \cite{GugV02} Given\/ $S$ and its associated structural relations\/
$\seqrel$, $\aseqrel$, $\parrel$ and\/ $\aparrel$, the following properties
hold, where\/ $a$, $b$, $c$ and\/ $d$ are distinct atom occurrences in\/ $S$:
\begin{description}
\item[$\sone$] None of\/ $\seqrel$, $\aseqrel$, $\parrel$ and\/ $\aparrel$ is
reflexive: $\lnot(a\seqrel a)$, $\lnot(a\aseqrel a)$, $\lnot(a\parrel a)$,
$\lnot(a\aparrel a)$.
\item[$\stwo$] One and only one among\/ $a\seqrel b$, $a\aseqrel b$, $a\parrel
b$ and\/ $a\aparrel b$ holds.
\item[$\sthree$] The relations\/ $\seqrel$ and\/ $\aseqrel$ are mutually inverse:
$a\seqrel b\Leftrightarrow b\aseqrel a$.
\item[$\sfour$] The relations\/ $\seqrel$ and\/ $\aseqrel$ are transitive:
$(a\seqrel b)\land(b\seqrel c)\Rightarrow a\seqrel c$ and\/ $(a\aseqrel
b)\land(b\aseqrel c)\Rightarrow a\aseqrel c$.
\item[$\sfive$] The relations\/ ${\parrel}$ and\/ ${\aparrel}$ are symmetric:
$a\parrel b\Leftrightarrow b\parrel a$ and\/ $a\aparrel b\Leftrightarrow
b\aparrel a$.
\item[$\ssix$] {\rm Triangular property}: for\/
$\sigma_1,\sigma_2,\sigma_3\in\{{\seqrel}\cup{\aseqrel},{\parrel},{\aparrel}\}$,
it holds
$$
(a\mathrel{\sigma_1}b)\land
(b\mathrel{\sigma_2}c)\land
(c\mathrel{\sigma_3}a)\Rightarrow
(\sigma_1=\sigma_2)\lor
(\sigma_2=\sigma_3)\lor
(\sigma_3=\sigma_1)
\quad.
$$
\item[$\sseven$] {\rm Square property}:
$$\openup1\jot
\begin{array}{rll}
\ssevenseq  \quad & 
(a\seqrel  b)\land
(a\seqrel  d)\land
(c\seqrel  d)\Rightarrow{}&
(a\seqrel  c)\lor
(b\seqrel  c)\lor
(b\seqrel  d)~\lor             \\
 & & 
(c\seqrel  a) \lor
(c\seqrel  b) \lor
(d\seqrel  b)\quad,                   \\
\ssevenpar \quad & 
(a\parrel  b)\land
(a\parrel  d)\land
(c\parrel  d)\Rightarrow{}&
(a\parrel  c)\lor
(b\parrel  c)\lor
(b\parrel  d)\quad,        \\
\ssevenapar \quad &
(a\aparrel b)\land
(a\aparrel d)\land
(c\aparrel d)\Rightarrow{}&
(a\aparrel c)\lor
(b\aparrel c)\lor
(b\aparrel d)\quad.       \\
\end{array}
$$
\end{description}
\end{thm}
\proof
See \cite{GugV02}.
\qed

\begin{figure}
$$
\begin{array}{rccccc}
\vcenter{\DiaSqProSeqOne} \Rightarrow & 
\vcenter{\DiaSqProSeqTwo} & \lor &
\vcenter{\DiaSqProSeqThree} & \lor &
\vcenter{\DiaSqProSeqFour} \\
& 
\vcenter{\DiaSqProSeqFive} & \lor &
\vcenter{\DiaSqProSeqSix} & \lor &
\vcenter{\DiaSqProSeqSeven} \\
\end{array}
$$
\caption{Square property for\/ $\seqrel$}
\label{figSqProSeq}
\end{figure}

\begin{figure}
$$
\begin{array}{rl}
\vcenter{\DiaSqProParOne} \Rightarrow \vcenter{\DiaSqProParTwo} \lor 
\vcenter{\DiaSqProParThree} \lor \vcenter{\DiaSqProParFour}
\end{array}
$$
\caption{Square property for\/ $\parrel$}
\label{figSqProPar}
\end{figure}

It is convenient to use graphical notations when reasoning
about relation webs. We use the following drawings 
$$\vcenter{
\xygraph{[]!{0;<2.3pc,0pc>:}
{a\strut}*[white]\cir<.5pc>{}-@{~>}@[seqcolor][r]
{b\strut}*[white]\cir<.5pc>{}                    }
}
\quad ,
\vcenter{
\xygraph{[]!{0;<2.3pc,0pc>:}
{a\strut}*[white]\cir<.5pc>{}-@{<~>}@[seqcolor][r]
{b\strut}*[white]\cir<.5pc>{}                     }
}
\quad ,
\vcenter{
\xygraph{[]!{0;<2.3pc,0pc>:}
{a\strut}*[white]\cir<.5pc>{}-@[parcolor][r]
{b\strut}*[white]\cir<.5pc>{}               }
}
\quad \hbox{\rm and} \quad
\vcenter{
\xygraph{[]!{0;<2.3pc,0pc>:}
{a\strut}*[white]\cir<.5pc>{}-@{~}@[aparcolor][r]
{b\strut}*[white]\cir<.5pc>{}                    }
}
$$
to represent the strutural relations: $a\seqrel b$ (and $b\aseqrel a$), 
$a\seqrel b$ or $a\aseqrel b$,
$a\parrel b$ and $a\aparrel b$, respectively.  
Dashed arrows represent negations of structural relations, e.g., 
$$
\vcenter{
\xygraph{[]!{0;<3pc,0pc>:}
{a\strut}*[white]\cir<.5pc>{}-@{--}@[antiparcolor][r]
{b\strut}*[white]\cir<.5pc>{}                     }
}
$$
means that $\lnot (a \parrel b)$.

Theorem~\ref{TheoStructRel} tells us that certain configurations
are absent in relation webs. 
The most interesting ones are those that derive from 
$\ssix$ (the triangular property) and $\sseven$ (the square property). 
The triangular property says that the following configuration is
absent in relation webs:
\begin{equation}
\label{eq:triangular}
\vcenter{
\xy
\xygraph{[]!{0;<-1.99185842pc,-1.15pc>:<-1.99185842pc,1.15pc>::} 
{c\strut}*[white]\cir<.5pc>{}(-  @{-}@[parcolor] [u]
{b\strut}*[white]\cir<.5pc>{},-  @{~}@[aparcolor][r]
{a\strut}*[white]\cir<.5pc>{} -@{<~>}@[seqcolor] [lu]
{ \strut}*[white]\cir<.5pc>{})                       }
\endxy
}
\end{equation}
For a structure with three atom occurrences, it is easy to 
check that the above configuration is absent: simply check all possible
structures that can be formed from the three atoms. For the more general
case, one needs to consider the cases where the atoms are nested
inside other substructures, e.g., as in $\pars{U\cons a, \seqs{V\cons b ; W\cons c}}$.
Induction on structures is needed in this case (see \cite{GugV02} for
details).

The square property for $\seqrel$ and $\parrel$ are presented 
graphically in Figure \ref{figSqProSeq} and 
Figure~\ref{figSqProPar}, respectively.
The square property for $\aparrel$ is similar to the one for $\parrel$.
From the square property, we can infer, for example, that there is no
structure containing the following configuration in its relation web
$$
\vcenter{
  \xymatrix@R=14pt@C=14pt{
     a \ar@[aparcolor]@{~}[d] \ar@[parcolor]@{-}[dr] \ar@[parcolor]@{-}[r]
     & b \ar@[aparcolor]@{~}[dl] \ar@[aparcolor]@{~}[d] \cr 
     c \ar@[parcolor]@{-}[r] & d \ar@[aparcolor]@{~}[u]
}} \quad .
$$

\begin{figure}
$$
\vcenter{
  \xymatrix@R=14pt@C=14pt{
     d \ar@[antiparcolor]@{--}[d] \ar@[parcolor]@{-}[dr] \ar@[antiparcolor]@{--}[r]
     & c \ar@[parcolor]@{-}[dl]  \cr 
     a \ar@[antiparcolor]@{--}[r] & b 
}} \quad 
\Rightarrow
\quad
\vcenter{
  \xymatrix@R=14pt@C=14pt{
     d \ar@[antiparcolor]@{--}[d] \ar@[parcolor]@{-}[dr] \ar@[antiparcolor]@{--}[r]
     & c \ar@[parcolor]@{-}[dl] \ar@[antiparcolor]@{--}[d] \cr 
     a \ar@[antiparcolor]@{--}[r] & b \ar@[antiparcolor]@{--}[u]
}} \quad 
$$
\caption{Inverse square property for $\parrel$}
\label{fig:invsq}
\end{figure}

A useful corollary of Theorem~\ref{TheoStructRel} is that the square property holds also for
some negated relations.
\begin{cor}
\label{CorInvSquare}
{\em Inverse square property.}
Given\/ $S$ and its associated structural relations\/
$\seqrel$, $\aseqrel$, $\parrel$ and\/ $\aparrel$, the following properties
hold, where\/ $a$, $b$, $c$ and\/ $d$ are distinct atom occurrences in\/ $S$:
\begin{enumerate}
\item  if $\lnot (a \parrel b)$, $\lnot (a \parrel d)$, $\lnot (c \parrel d)$,
$a \parrel c$ and $b \parrel d$ then $\lnot (b \parrel c)$,
\item  if $\lnot (a \aparrel b)$, $\lnot (a \aparrel d)$, $\lnot (c \aparrel d)$,
$a \aparrel c$ and $b \aparrel d$ then $\lnot (b \aparrel c)$.
\end{enumerate}
\end{cor}
\proof
We show the case for the $\parrel$ relation, the other case is proved analogously.
Suppose that  $\lnot (a \parrel b)$, $\lnot (a \parrel d)$, $\lnot (c \parrel d)$,
$a \parrel c$ and $b \parrel d$ hold, but $b \parrel c$. Then applying the square
property for $\parrel$ in Theorem~\ref{TheoStructRel}, it must be the case that one of the 
following relations holds: $a \parrel b$, $a \parrel d$, $c \parrel d$. But either one
of them would result in a contradiction with the assumptions. 
\qed
Figure~\ref{fig:invsq} shows the graphical representation of the inverse square property for 
the $\parrel$ relation.
Note that the inverse square property for the $\seqrel$ relation, i.e.,
``if $\lnot (a \seqrel b)$, $\lnot (a \seqrel d)$, $\lnot (c \seqrel d)$,
$a \seqrel c$ and $b \seqrel d$ then $\lnot (b \seqrel c)$'' does not hold.
Consider for instance the structure $\seqs{b ; \pars{\seqs{a ; c}, d}}$.
Its relation web is the following:
$$
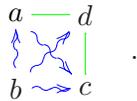
\vcenter{
  \xymatrix@R=14pt@C=14pt{
     a \ar@[seqcolor]@{<~}[d] \ar@[seqcolor]@{~>}[dr] \ar@[parcolor]@{-}[r]
     & d \ar@[seqcolor]@{<~}[dl] \ar@[parcolor]@{-}[d] \cr 
     b \ar@[seqcolor]@{~>}[r] & c 
}} \quad .
$$
We have in the web $\lnot (a \seqrel b)$, $\lnot (a \seqrel d)$, $\lnot (c \seqrel d)$,
$a \seqrel c$ and $b \seqrel d$ but $b \seqrel c$.

The conditions $\sone$ - $\sseven$ in Theorem~\ref{TheoStructRel} turn out to be sufficient 
for characterizing structures, that is, every {\em web candidate} (i.e., a graph with edges labelled 
with the relations $\seqrel$, $\parrel$ and $\aparrel$) which satisfies the conditions 
is a relation web (hence a structure). This is proved in \cite{GugV02} where a non-deterministic
algorithm is given to decide whether a web candidate is a relation web. 
We shall not go into the details of the general algorithm and its correctness proof since they are not needed in
establishing the main results of this paper. We illustrate here briefly how one can recover
a structure from a given relation web using the algorithm. The algorithm works by identifying substructures
in the relation web, starting from the smallest ones. At each step of the algorithm,
partitions are formed on the web such that each partition corresponds to a valid (sub)structure.
Given a relation web $\wbS$, the steps to recover the structure corresponding to the
web is as follows:
\begin{description}
\item[First step] Form a partition $\{a \}$ for each atom occurrence $a$ in the web. 
\item[Iterative step] 
Let $\mu$ and $\nu$ be two distinct partitions (which are valid substructures) 
such that all atom occurrences in $\mu$ and $\nu$ are in the same relations, and
for all $a \in \mu$, $b \in \nu$ and $c \in \occ S \setminus (\mu \cup \nu)$, 
$a \seqrel c$ (respectively, $a \parrel c$ and $a \aparrel c$) 
iff $b \seqrel c$ (respectively, $b \parrel c$ and $b \aparrel c$). 
Merge the partitions $\mu$ and $\nu$. Let $U$ be the corresponding structure of 
$\mu$ and $V$ be the corresponding structure of $\nu$. If $\mu$ and $\nu$
are related by $\parrel$ (respectively, $\seqrel$ and $\aparrel$) 
then the corresponding structure for the merged partition is 
$\pars{U, V}$ (respectively, $\seqs{U ; V}$ and $\aprs{U, V}$).
Repeat this step until no merging is possible. The structure corresponding to
the remaining partition (which should be the relation web $\wbS$ itself) is
the structure of the relation web. 
\end{description} 
Consider for instance the following web:
$$
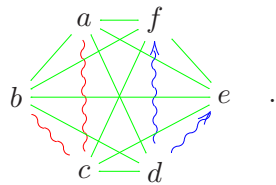
\vcenter{
\xymatrix@R=14pt@C=14pt{
   & a \ar@[parcolor]@{-}[dl]  \ar@[aparcolor]@{~}[dd] \ar@[parcolor]@{-}[ddr]
       \ar@[parcolor]@{-}[drr] \ar@[parcolor]@{-}[r]
   & f
       \ar@[parcolor]@{-}[dll] \ar@[parcolor]@{-}[ddl] \ar@[seqcolor]@{<~}[dd]
       \ar@[parcolor]@{-}[dr]
   & \\
b  \ar@[aparcolor]@{~}[dr] \ar@[parcolor]@{-}[drr] \ar@[parcolor]@{-}[rrr]  
   &  &
   & e
     \ar@[parcolor]@{-}[dll]  \ar@[seqcolor]@{<~}[dl]
   \\
   & c 
     \ar@[parcolor]@{-}[r]
   & d &   
}}
\quad .
$$
The intermediate partitions produced in the execution of the above algorithm is as follows,
where we represent each partition with its corresponding structure:
\begin{enumerate}
\item $a$, $b$, $c$, $d$, $e$, $f$.
\item $\pars{a, b}$,  $c$, $d$, $e$, $f$.
\item $\pars{a, b}$, $c$, $d$, $\pars{e,f}$.
\item $\aprs{\pars{a,b}, c}$, $d$, $\pars{e,f}$.
\item  $\aprs{\pars{a,b}, c}$, $\seqs{d ; \pars{e,f}}$.
\item $\pars{\aprs{\pars{a,b}, c}, \seqs{d ; \pars{e,f}}}$.
\end{enumerate}
The above relation web therefore corresponds to the structure
$\pars{\aprs{\pars{a,b}, c}, \seqs{d ; \pars{e,f}}}.$

The following theorem states that relation webs are a canonical representation
of structures.

\begin{thm}
\label{TheoTrace} 
\cite{GugV02} Two structures are equivalent if and only if they have 
the same relation web.
\end{thm}

\section{Shallow Systems}
\label{sec:shallow}

In the definition of shallow systems to follow, we shall try to be as general as possible
but should also be careful to rule out certain inconsistent rules. An obvious
case would be rules which would allow one to prove $A \lpar B \limp A\ltens B$, for example.
Two conditions are enforced on the definition of candidate shallow systems: a shallow system
should have a notion of cut and cut-elimination property, and further, the inference
rules should respect a certain ordering of logical strength of structural relations.
The latter can be seen as a sort of ``subformula'' property. We observe the following
ordering on the structural relations: $\parrel ~ \prec ~ \seqrel ~ \prec ~ \aparrel$
and $\parrel ~ \prec ~ \aseqrel ~ \prec ~ \aparrel$.

\begin{defi}
\label{DefStructOrder}
Let $R$ and $T$ be two structures such that $R \not = T$ and $\occ R = \occ T$.
The relation $R \prec T$ holds if for all atom occurrences $a$ and $b$ in $R$,
the following hold:
\begin{enumerate}
\item if $a \parrel_T b$ then $a \parrel_R b$,
\item if $a \aparrel_T b$ then either one of the following holds: 
$a \seqrel_R b$, $a \aseqrel_R b$, $a \aparrel_R b$ or $a \parrel_R b$,
\item if $a \seqrel_T b$ then either $a \seqrel_R b$ or $a \parrel_R b$,
\item if $a \aseqrel_T b$ then either $a \aseqrel_R b$ or $a \parrel_R b$.
\end{enumerate}

\end{defi}

We shall now formalize what we mean by shallow inference rules and shallow systems.
For this we define formally the notion of structure schemes, that is, structures containing
variables. Structure scheme is then used in defining rule scheme and the depth
of a rule scheme.

\begin{defi}
\label{DefGenStruct}
We enrich the language of structures  
in Definition~\ref{DefStructures} with a denumerable infinite set of structure variables. Structure variables
are denoted with $A$, $B$ and $C$. The extended language of structures is defined 
as follows:
$$
S \grammareq
             a                                        \mid
             A                                        \mid
             \un                                      \mid             
             \pars{\,\underbrace{S,\dots,S}_{{}>0}\,} \mid
             \aprs{\,\underbrace{S,\dots,S}_{{}>0}\,} \mid
             \seqs{\,\underbrace{S;\dots;S}_{{}>0}\,} \mid
             \neg S                                       \quadcm
$$
where $a$ is an atom and $A$ is a structure variable. Structure
variables representing atomic structures will also be denoted
with $x$, $y$ and $z$.  
The definitions that apply to structures without variables 
apply also to structures with variables.
In particular, $\occ S$ now denotes the set of occurences of
atoms and variables in the structure $S$. Definition~\ref{DefStructOrder} on
the ordering of structures extends straightforwardly to structures
with variables. The set of structure
variables in $S$ is denoted by $\vars S$.  A {\em ground} structure is 
a structure that does not contain any variable. A structure with
variables can be {\em instantiated} by replacing the variables with other
structures. 
\end{defi}

In the following definition of shallow systems, only the logical rules (i.e., the non-interaction
rules) are required to be shallow, while the interaction rules are those of $\BV$. 
Moreover, since the systems that are of interest will be those with 
cut-elimination property, we always refer to cut-free systems when we talk 
about shallow systems. 

\begin{defi}
\label{def:shallow-rules}
The {\em depth} of a structure $S$ is defined as follows:
$$\depth S = \max{\{\depth S'\ec \mid S'\cons{R}=S \hbox{ and } R \in \occ S \}}$$
An inference rule $\vcenter{\infnote{\rho}{R}{T}{}}$ is {\em shallow} if 
$R$ and $T$ are non-unit structures such that $R \prec T$ and the variables in each of $R$ and $T$ 
are pairwise distinct. 
The depth of $\rho$ is defined as $\max{(\depth{R}, \depth{T})}$.
\end{defi}

Notice that the above definition of shallow inference rules
excludes the interaction, contraction and weakening rules.
An example of a shallow rule is the following rule 
$$\vcenter{\inf{\rho}{\pars{A, B, \aprs{C,C'}}}
                     {\pars{A, \aprs{\pars{B,C},C'}}}}.$$
It is a shallow rule with depth 3, which is the depth of the structure in the premise.

\begin{defi}
\label{def:shallow-systems}
A shallow system $\sysS$ consists of the rules 
$\{\intr, \unr\}$ and a set of shallow rules $\sysS'$ such that 
there exists an $n$, the {\em depth} of $\sysS$, such that for every rule $\rho \in \sysS'$, 
$\depth{\rho} \leq n$.
\end{defi}

In proving general properties of shallow rules, we shall use the relation webs semantics
to characterize their operations on structures. 
An application of a non-interaction rule, seen at the relation webs level, modifies 
structural relations between atom occurrences in a structure. 
The square and triangular properties can be used to infer the effects on the overall 
structural relations in the structure. In addition, we also make use of the fact that 
there are some structural relations that are not modified by a shallow rule, 
if they happen to belong to a substructure that is nested deeper
than the depth of the shallow rule. This is stated in the following lemma.

\begin{defi}
Let $\vcenter{\inf{\rho}{R}{T}}$ be an instance of a rule $\rho$. The structural
relation $a\ \sigma_R \ b$, where $a,b \in \occ R$ and 
$\sigma \in \{ \parrel, \seqrel, \aparrel \}$,
is {\em preserved} by $\rho$ if and only if $a,b \in \occ T$ and $a\ \sigma_T \ b$.
\end{defi}

\begin{lem}
\label{LemmaSubStructure}
Let $\vcenter{\inf{\rho}{R}{T}}$ be an instance of a shallow rule $\rho$ of depth $n$
and $P$ be a structure occuring in $R$ at depth $m$. If $m > n$ then all the structural relations
in $P$ are preserved by $\rho$.
\end{lem}
\proof
The proof follows straightforwardly from the definition of shallow rules and relation webs.
\qed

\section{The Need for Deep Inference}
\label{sec:deep}

A main part of the proof of the necessity of deep-inference for $\BV$
is the construction of the class of provable structures which will serve 
as the counterexample to the completeness of any given shallow
candidate system for $\BV$. This class of structures is generated from the structure 
$$S_0 = \pars{\seqs{\pars{a,b};c}, \seqs{\neg a;\pars{\neg b,\neg c}}}.$$ 
It is done by merging recursively variants of $S_0$ with the innermost 
par-substructures of the original structure. 
The formal definition of this merging process is as follows.

\begin{defi} 
A structure $R$ is called {\em flat} if all the structural relations
between atom occurrences in $R$ are of the same type. The structure $\pars{a_1, \dots, a_n}$
is a {\em flat par structure}. Likewise, $\aprs{a_1,\dots,a_n}$ and $\seqs{a_1;\dots;a_n}$
are a {\em flat copar structure} and a {\em flat seq structure}, respectively. A unit or 
an atomic structure can be seen as a flat par structure, flat copar structure or flat 
seq structure.
\end{defi}

\begin{defi}
\label{DefFractalStruct}
Let ${\bf\rm C}$ be an infinite 
set of positive atoms and let ${\bf\rm N}^*$ be the set of finite sequences of natural 
numbers. Elements of ${\bf\rm N}^*$ are denoted by $u$, $v$, $w$ and $z$. 
Concatenation of two sequences of natural number $u$ and $v$ is denoted by $u.v$. 
The set of indexed atoms is defined as $\CN = \{ a_u \mid a \in {\bf\rm C},\  
u \in {\bf\rm N}^* \}$. Two
atoms with different names or indexes are considered different. 
The following functions generate a class of structures:
$$
\begin{matrix}
\alpha_0(u,R,T) & = & \SM{a_u}{b_u}{c_u}{R}{T} \hfill \cr
\alpha_n(u,R,T) & = & 
       \pars{\seqs{\alpha_{n-1}(u.0, a_u, \pars{b_u,R});c_u}, 
       \seqs{\neg a_u;\alpha_{n-1}(u.1, \neg b_u, \pars{\neg c_u, T})}}, 
       \hbox{ {\rm if} $n > 0$}, 
\end{matrix}
$$
where $u \in {\bf\rm N}^*$, $a\not = b \not = c \not = a$ and $R,T$ are 
flat par structures. A structure of the form $\alpha_n(u,R,T)$ is called an 
$\alpha_n$-structure. We denote with $S_n$ the structure $\alpha_n(0,\un,\un)$.
\end{defi}

Note that the use of indices in naming atoms in the above definition is
just to guarantee the pairwise-distinctness of atoms in the generated
structures. In the following discussions, we shall often drop the indices
when it is understood from context that the atoms under consideration
are pairwise distinct.

Recall from the proof in Section~\ref{sec:overview} that proof search for the structure
$S_0$ (and, as we shall see later, any other $S_n$) must start with certain innermost redexes 
in the structure. A way to show the necessity of deep inference is to show that for each $n$, 
any rules that modify the structural relations in $S_n$ but leave the innermost redexes unchanged 
would result in an unprovable structure. However, this would leave us with infinite
possible inference rules to consider. We therefore need a finite characterization of inference
rules, which is complete for the class of structures under consideration. 
We use the relation web representation for this purpose. 
Recall that a (shallow) inference rule, applied to a structure, modifies structural relations
between atom occurrences in the structure. 
For instance, the following rule
$$
\inf{\rho}{\pars{R, \aprs{T,U}}}{\aprs{\pars{R,T}, U}}
$$
can be seen, at the relation web level, as the instruction (reading
the rule bottom up): change all the structural relations between the
atom occurrences in $R$ and $U$ to $\aparrel$. 
Of course, one cannot arbitrarily modify the structural relations in
a relation web, since the resulting web may not correspond to a valid
structure. The characterization of structures in Theorem~\ref{TheoStructRel}
is used to rule out certain invalid operations on the relation webs (i.e.,
by showing that certain forbidden configurations would result from
them). However, this characterization alone is not enough
for showing the necessity of deep inference, since they deal with
structures in general, while we require as well characterization of
provable structures. The following lemma states that there are  additional
configurations that are absent from provable structures.

\begin{lem}
\label{LemmaNoCross}
Let $S$ be a structure consisting of pairwise distinct atoms. 
If $S$ is provable in $\BV$ then the relation web of $S$ does not contain 
any of the following configurations:
$$
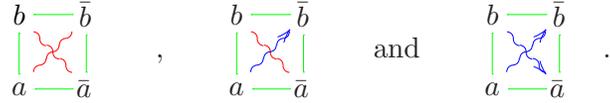
\DiaForbiddenConf
$$
\end{lem}
\proof
Suppose that $S$ is provable in $\BV$ but $\wb_S$ contains such configurations,
i.e., there exists atoms $a,\neg a, b, \neg b \in \occ S$ that are interrelated
as in one of the configurations above. 
Take any proof $\vcenter{ \ProofDia{S}{\Pi}{\BV}{d}{2} }$. By repeated applications
of Proposition~\ref{PropProofSubst} we can remove all the atom occurrences from 
$\Pi$, except for the atoms $\{ a,b,\neg a, \neg b\}$, and obtain the proof
$\vcenter{\ProofDia{S'}{\Pi'}{\BV}{d}{2}}$. The relation web of $S'$ must be in one of 
the configurations above, which correspond to the structures 
$\pars{\aprs{a,\neg b}, \aprs{\neg a, b}}$,
$\pars{\seqs{a; \neg b}, \aprs{\neg a, b}}$ and
$\pars{\seqs{a; \neg b}, \seqs{b;\neg a}}$, respectively. 
As can be easily checked, none of these structures are provable in $\BV$,
which contradicts the fact that $S'$ is provable.
\qed

We illustrate here how the characterizations of (provable) structures can be
used to show that $S_0$ can be proved only by modifying the innermost redexes first.
This will be shown using in particular the square and triangular properties and the results from 
Lemma~\ref{LemmaNoCross}, Lemma~\ref{LemmaSubStructure} and 
Definition~\ref{DefStructOrder}.
The relation web of $S_0$ (the indexes are omitted) is the following:
$$
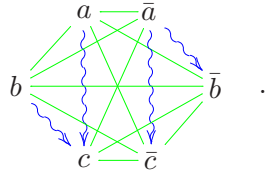
\vcenter{\DiaHexagonConf} \quad .
$$
Suppose, for example, there is a proof of $S_0$ that starts with a rule $\rho$ of depth 1.
Then by Lemma~\ref{LemmaSubStructure}, the structural relations $a \parrel b$
and $\neg b \parrel \neg c$ must be preserved by $\rho$ since they correspond to
substructures at depth 2. Assuming the application of $\rho$ is non-trivial, some other 
structural relations must be changed. The changing, however, must preserve provability, 
which means  the par relations between dual atoms must also be preserved. 
Given these constraints, any other changes to the structural 
relations will either result in a violation of a certain structural property or 
destroy the provability. 
For example, if we change $a \parrel \neg c$
to $a \aparrel \neg c$, the triangular and square properties will be violated:
$$\vcenter
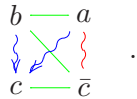
{
  \xymatrix@R=14pt@C=14pt{
     b \ar@[seqcolor]@{~>}[d] \ar@[parcolor]@{-}[dr] \ar@[parcolor]@{-}[r]
     & a \ar@[seqcolor]@{~>}[dl] \ar@[aparcolor]@{~}[d] \cr
     c \ar@[parcolor]@{-}[r] & \neg c
}} \quad .
$$
This idea can be generalised to the case 
$$S_n = \pars{\seqs{\pars{P\cons{a}, Q\cons{b}};c}, 
             \seqs{\neg a;\pars{P'\cons{\neg b}, Q'\cons{\neg c}}}}.$$ 
If we preserve all the
structural relations inside $\pars{P\cons{a}, Q\cons{b}}$ and
$\pars{P'\cons{\neg b}, Q'\cons{\neg c}}$, then all other structural relations cannot be changed
without changing the provability of $S_n$. This idea is formalized in 
Lemma~\ref{LemmaSnConfig} and Lemma~\ref{LemmaAlphaOne}.

\begin{lem}
\label{LemmaDrvAlpha}
For every flat par structures $R$ and $T$, and for every
$u \in {\bf\rm N}^{*}$ there exists a derivation 
from $\pars{R,T}$ to $\alpha_n(u,R,T)$ in $\BV$.
\end{lem}
\proof

{\em Base Case:}  $\alpha_0(u,R,T) = \SM{a_u}{b_u}{c_u}{R}{T}.$ 
The derivation needed for this case is given in Figure~\ref{fig:param-S0} (modulo some
renaming of atoms).

{\em Inductive Case:}
Assume that  $\vcenter{\drv{\alpha_{i}(v,U,V)}{\pars{U,V}}{}{\BV}{d}}$ for all 
$i<n$, for all $v \in {\bf\rm N}^*$ and for all flat par structures $U,V$. Then
$$
\xy
\xygraph{[]!{0;<2pc,0pc>:}
{
  \pars{R,T}
}-@2^<>(.5){} 
  _<>(.5){\Delta_0} [d]
{ 
        \pars{\seqs{\pars{a_u, b_u,R}; c_u}, 
              \seqs{\neg a_u; \pars{\neg b_u, \neg c_u, T}}
              }
}-@2^<>(.5){}
 _<>(.5){\Delta_1}[d]
{
  \pars{\seqs{\pars{a_u, b_u,R}; c_u}, 
        \seqs{\neg a_u; \alpha_{n-1}(u.1, \neg b_u, \pars{\neg c_u, T})}
              }
}-@2^<>(.5){}
 _<>(.5){\Delta_2}[d]
{
\alpha_n(u, R, T) = \pars{\seqs{\alpha_{n-1}(u.0, a_u, \pars{b_u,R}); c_u}, 
              \seqs{\neg a_u; \alpha_{n-1}(u.1, \neg b_u, \pars{\neg c_u, T})}
              }
}                                  }
\endxy
$$
where $\Delta_0$ is the derivation constructed in the base case, and 
$\Delta_1$ and $\Delta_2$ are derivations obtained from induction hypothesis.
\qed

\begin{lem}
\label{LemmaSnProvable}
For every $n$, $S_n$ is provable in $\BV$.
\end{lem}
\begin{proof}
Follows immediately from Lemma~\ref{LemmaDrvAlpha}, since 
$S_n = \alpha_n(0,\un,\un)$.
\end{proof}

\begin{lem}
\label{LemmaNoDualStrInSn}
There are no substructures of the form 
$\pars{R, \overline{R}}$ in $S_n$.
\end{lem}
\proof

{\em Base Case:}
$S_0 = \alpha_0(0,\un,\un) = \pars{\seqs{\pars{a_0,b_0};c_0}, 
                                   \seqs{\neg a_0;\pars{\neg b_0, \neg c_0}}}$, obvious.

{\em Inductive Case:} 
Assume that $S_{n-1} = \alpha_{n-1}(0,\un,\un)$ has no substructures
of the form $\pars{R, \neg R}$. 
The change of index from $\alpha_{n-1}(0,\un,\un)$ to $\alpha_{n-1}(u,\un,\un)$ 
does not affect the form of the structure, since all the indexed atoms are 
distinct by definition. The addition of new atoms to 
$\alpha_{n-1}(u,\un,\un)$, i.e., $\alpha_{n-1}(u,U,V)$, 
does not introduce any new dual substructures, provided that $U$ 
and $V$ have no dual atoms and consist of different atoms than
$\alpha_{n-1}(u, \un, \un)$. Therefore the structure
$$S_n = \alpha_n(0,\un,\un) = \pars{ \seqs{\alpha_{n-1}(0.0,a_0,b_0);c_0}, 
              \seqs{\neg a_0;\alpha_{n-1}(0.1,\neg b_0, \neg c_0) }}$$
does not contain any substructures of the form $\pars{R,\neg R}$ either.
\qed

\begin{lem}
\label{LemmaDepthOfAlphaZero}
Every occurrence of $\alpha_0$-structure in
$\alpha_n(u,R,T)$ is at depth $2n$.
\end{lem}
\proof
The base case is obvious. For the inductive case, assume that every
occurrence of $\alpha_0$-structure in $\alpha_{n-1}(w,P,Q)$ is at depth $2(n-1)$. 
Since $$\alpha_n(u,R,T) = \pars{\seqs{\alpha_{n-1}(u.0, a_u, \pars{b_u,R});c_u}, 
       \seqs{\neg a_u;\alpha_{n-1}(u.1, \neg b_u, \pars{\neg c_u, T})}},$$ the depth
of each $\alpha_0$-structure in $\alpha_n(u,R,T)$ is $2(n-1)+2=2n$.
\qed

The next three lemmas are crucial to the proof by contradiction in the main theorem 
on the necessity of deep inference.

\begin{lem}
\label{LemmaSnConfig}
Let 
$\vcenter{
\ProofDia{\hbox{\infnote {\hbox to0pt{\hss$\rho$}} {R} {T} {}} } {} {\sysS}{d}{2.3}
}$ be a proof in a shallow system $\sysS$ where $\rho$ is a shallow rule. If $\sysS$ is equivalent to $\BV$ 
and the following conditions hold:
\begin{description}
\item[(a)] $R$ consists of pairwise distinct atoms,
\item[(b)] $\wb_R$ contains the following configuration:
$$
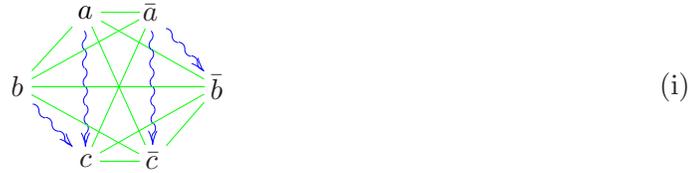
\vcenter{\DiaHexagonConf}
\eqno({\rm i})
$$
for some pairwise distinct atoms $a$, $b$ and $c$, 
\item[(c)]$a \parrel_T b$ and $\neg b \parrel_T \neg c$, 
\end{description}
then the configuration (i) is preserved by $\rho$.
\end{lem}

\proof
Since $T$ is provable, the par links between dual atoms in $R$ must be preserved by $\rho$.
This, together with conditions (a)-(c), give us the following configuration in $\wb_T$:
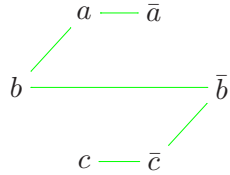
$$
\vcenter{
\xymatrix@R=14pt@C=14pt{
   & a \ar@[parcolor]@{-}[dl]  
       \ar@[parcolor]@{-}[r]
   & \neg a 
   & \\
b  \ar@[parcolor]@{-}[rrr]  
   &  &
   & \neg b 
     \ar@[parcolor]@{-}[dl]
   \\
   & c 
     \ar@[parcolor]@{-}[r]
   & \neg c &   
}}
$$
We assume above that $\sysS$ is equivalent to $\BV$. By Definition~\ref{def:shallow-systems} and
Definition~\ref{def:shallow-rules}, we have that $R \prec T.$
This means, by Definition~\ref{DefStructOrder}, that if $x \seqrel_R y$ or $x \aparrel_R y$ then it cannot be the case that
$x \parrel_T y$. Therefore it must hold in $\wb_T$:
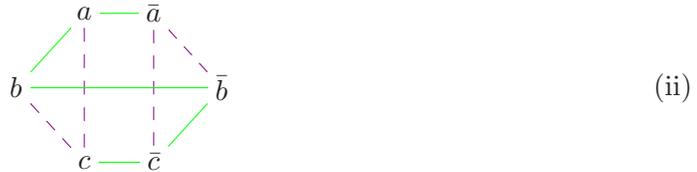
$$
\vcenter{
\xymatrix@R=14pt@C=14pt{
   & a \ar@[parcolor]@{-}[dl]  \ar@[antiparcolor]@{--}[dd] 
       \ar@[parcolor]@{-}[r]
   & \neg a 
       \ar@[antiparcolor]@{--}[dd]
       \ar@[antiparcolor]@{--}[dr]
   & \\
b  \ar@[antiparcolor]@{--}[dr] \ar@[parcolor]@{-}[rrr]  
   &  &
   & \neg b 
     \ar@[parcolor]@{-}[dl]
   \\
   & c 
     \ar@[parcolor]@{-}[r]
   & \neg c &   
}}
\eqno({\rm ii})
$$
(Recall that the dotted line indicates the negation of the structural relation it
represents, in this case, the par link.) 
We first show that all par relations $x\parrel_R y$ in (i) 
must be preserved, and then, based on this result we prove that 
any seq relations $x \seqrel_R y$ in (i) must also
be preserved. 

\begin{description}
\item[1.] $a \parrel_T \neg a$, $b \parrel_T \neg b$, $c \parrel_T \neg c$,
          $a \parrel_T b$ and $\neg b \parrel_T \neg c$, see diagram (ii).
\item[2.] $a \parrel_T \neg b$. Suppose that $\lnot(a \parrel_T \neg b )$.
Adding this information to (ii) we have in $\wb_T$:
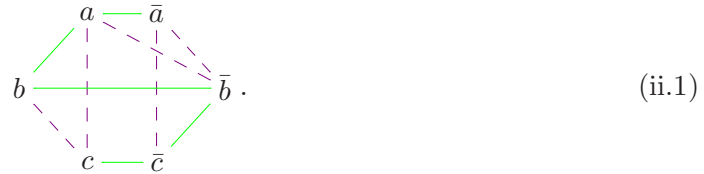
$$
\vcenter{
\xymatrix@R=14pt@C=14pt{
   & a \ar@[parcolor]@{-}[dl]  \ar@[antiparcolor]@{--}[dd] 
       \ar@[antiparcolor]@{--}[drr] \ar@[parcolor]@{-}[r]
   & \neg a 
       \ar@[antiparcolor]@{--}[dd]
       \ar@[antiparcolor]@{--}[dr]
   & \\
b  \ar@[antiparcolor]@{--}[dr] \ar@[parcolor]@{-}[rrr]  
   &  &
   & \neg b 
     \ar@[parcolor]@{-}[dl]
   \\
   & c 
     \ar@[parcolor]@{-}[r]
   & \neg c &   
}}.
\eqno({\rm ii.1})
$$  
Take the atoms $a$, $\neg a$, $\neg b$ and $\neg c$. 
By the inverse square property (Corollary~\ref{CorInvSquare}), we infer
that $\lnot(a \parrel_T \neg c)$, as shown in the following figure:
$$
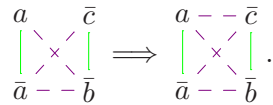
\vcenter{
  \xymatrix@R=14pt@C=14pt{
     a \ar@[parcolor]@{-}[d] \ar@[antiparcolor]@{--}[dr]
     & \neg c \ar@[antiparcolor]@{--}[dl] \ar@[parcolor]@{-}[d] \cr
     \neg a \ar@[antiparcolor]@{--}[r] & \neg b
  }
}
\Longrightarrow
\vcenter{
  \xymatrix@R=14pt@C=14pt{
     a \ar@[parcolor]@{-}[d] \ar@[antiparcolor]@{--}[dr]
       \ar@[antiparcolor]@{--}[r]
     & \neg c \ar@[antiparcolor]@{--}[dl] \ar@[parcolor]@{-}[d] \cr
     \neg a \ar@[antiparcolor]@{--}[r] & \neg b
  }
}.
$$
Now we combine this information with (ii.1):
$$
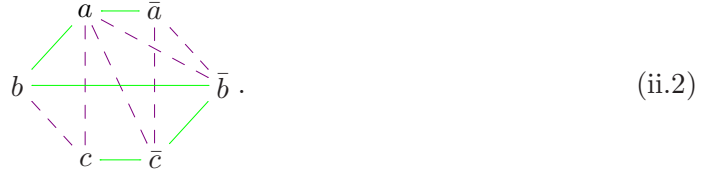
\vcenter{
\xymatrix@R=14pt@C=14pt{
   & a \ar@[parcolor]@{-}[dl]  \ar@[antiparcolor]@{--}[dd] 
       \ar@[antiparcolor]@{--}[drr] \ar@[parcolor]@{-}[r]
       \ar@[antiparcolor]@{--}[ddr]
   & \neg a 
       \ar@[antiparcolor]@{--}[dd]
       \ar@[antiparcolor]@{--}[dr]
   & \\
b  \ar@[antiparcolor]@{--}[dr] \ar@[parcolor]@{-}[rrr]  
   &  &
   & \neg b 
     \ar@[parcolor]@{-}[dl]
   \\
   & c 
     \ar@[parcolor]@{-}[r]
   & \neg c &   
}}.
\eqno({\rm ii.2})
$$
Take the atoms $b$, $\neg b$, $a$ and $c$. By the square property (Theorem~\ref{TheoStructRel}),
$b \parrel_T \neg c$ as shown below: 
$$
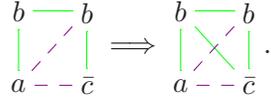
\vcenter{
  \xymatrix@R=14pt@C=14pt{
     b \ar@[parcolor]@{-}[d] \ar@[parcolor]@{-}[r]
     & \neg b \ar@[antiparcolor]@{--}[dl] \ar@[parcolor]@{-}[d] \cr
     a \ar@[antiparcolor]@{--}[r] & \neg c
  }
}
\Longrightarrow
\vcenter{
  \xymatrix@R=14pt@C=14pt{
     b \ar@[parcolor]@{-}[d] \ar@[parcolor]@{-}[r]
       \ar@[parcolor]@{-}[dr]
     & \neg b \ar@[antiparcolor]@{--}[dl] \ar@[parcolor]@{-}[d] \cr
     a \ar@[antiparcolor]@{--}[r] & \neg c
  }
}.
$$
This together with (ii.2) yield:
$$
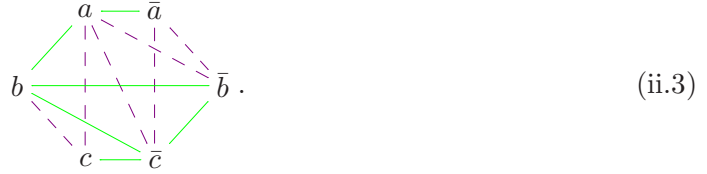
\vcenter{
\xymatrix@R=14pt@C=14pt{
   & a \ar@[parcolor]@{-}[dl]  \ar@[antiparcolor]@{--}[dd] 
       \ar@[antiparcolor]@{--}[drr] \ar@[parcolor]@{-}[r]
       \ar@[antiparcolor]@{--}[ddr]
   & \neg a 
       \ar@[antiparcolor]@{--}[dd]
       \ar@[antiparcolor]@{--}[dr]
   & \\
b  \ar@[antiparcolor]@{--}[dr] \ar@[parcolor]@{-}[rrr]  
   \ar@[parcolor]@{-}[drr]
   &  &
   & \neg b 
     \ar@[parcolor]@{-}[dl]
   \\
   & c 
     \ar@[parcolor]@{-}[r]
   & \neg c &   
}}.
\eqno({\rm ii.3})
$$
But if this is the case, $T$ cannot be a valid structure, because
$
\vcenter{
  \xymatrix@R=14pt@C=14pt{
     b \ar@[parcolor]@{-}[d] \ar@[parcolor]@{-}[r]
       \ar@[antiparcolor]@{--}[dr]
     & \neg c \ar@[antiparcolor]@{--}[dl] \ar@[parcolor]@{-}[d] \cr
     a \ar@[antiparcolor]@{--}[r] & c
  }
}
$ 
violates the square property. Therefore $a\parrel \neg b$ must hold in $T$. 

\item[3.] $a \parrel \neg c$ in $T$. Suppose that $\lnot(a \parrel \neg c)$,
following a similar reasoning as in Case 2, we have (s.p. stands for the (inverse) square
property):
$$
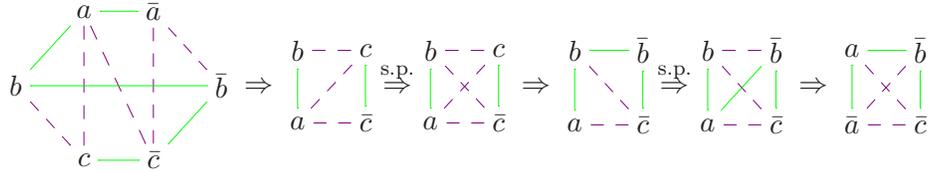
\vcenter{
\xymatrix@R=14pt@C=14pt{
   & a \ar@[parcolor]@{-}[dl]  \ar@[antiparcolor]@{--}[dd] 
       \ar@[parcolor]@{-}[r]
       \ar@[antiparcolor]@{--}[ddr]
   & \neg a 
       \ar@[antiparcolor]@{--}[dd]
       \ar@[antiparcolor]@{--}[dr]
   & \\
b  \ar@[antiparcolor]@{--}[dr] \ar@[parcolor]@{-}[rrr]  
   &  &
   & \neg b 
     \ar@[parcolor]@{-}[dl]
   \\
   & c 
     \ar@[parcolor]@{-}[r]
   & \neg c &   
}}
\Rightarrow
\vcenter{
\xymatrix@R=14pt@C=14pt{
  b \ar@[parcolor]@{-}[d] \ar@[antiparcolor]@{--}[r] & 
  c \ar@[antiparcolor]@{--}[dl] \ar@[parcolor]@{-}[d] \cr
  a \ar@[antiparcolor]@{--}[r] & \neg c
}
}
{\buildrel {\rm s.p.} \over \Rightarrow}
\vcenter{
\xymatrix@R=14pt@C=14pt{
  b \ar@[parcolor]@{-}[d] \ar@[antiparcolor]@{--}[r] 
    \ar@[antiparcolor]@{--}[dr] & 
  c \ar@[antiparcolor]@{--}[dl] \ar@[parcolor]@{-}[d] \cr
  a \ar@[antiparcolor]@{--}[r] & \neg c
}
}
\Rightarrow
\vcenter{
\xymatrix@R=14pt@C=14pt{
  b \ar@[parcolor]@{-}[d] \ar@[parcolor]@{-}[r] 
    \ar@[antiparcolor]@{--}[dr] & 
  \neg b \ar@[parcolor]@{-}[d] \cr
  a \ar@[antiparcolor]@{--}[r] & \neg c
}
}
{\buildrel {\rm s.p.} \over \Rightarrow}
\vcenter{
\xymatrix@R=14pt@C=14pt{
  b \ar@[parcolor]@{-}[d] \ar@[antiparcolor]@{--}[r] 
    \ar@[antiparcolor]@{--}[dr] & 
  \neg b \ar@[parcolor]@{-}[dl] \ar@[parcolor]@{-}[d] \cr
  a \ar@[antiparcolor]@{--}[r] & \neg c
}
}
\Rightarrow
\vcenter{
\xymatrix@R=14pt@C=14pt{
  a \ar@[parcolor]@{-}[d] \ar@[parcolor]@{-}[r] 
    \ar@[antiparcolor]@{--}[dr] & 
  \neg b \ar@[antiparcolor]@{--}[dl] \ar@[parcolor]@{-}[d] \cr
  \neg a \ar@[antiparcolor]@{--}[r] & \neg c
}
}
$$

\item[4.] $b \parrel_T \neg a$. Suppose $\lnot(b \parrel_T \neg a)$.
It follows from the results established in cases 1-3 that in $\wb_T$: 
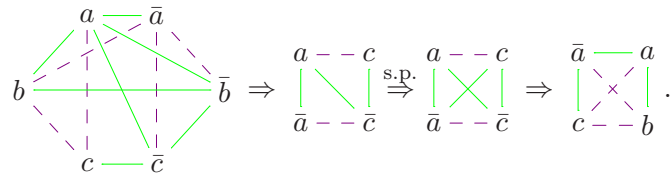
$$
\vcenter{
\xymatrix@R=14pt@C=14pt{
   & a \ar@[parcolor]@{-}[dl]  \ar@[antiparcolor]@{--}[dd] 
       \ar@[parcolor]@{-}[r]  \ar@[parcolor]@{-}[drr]
       \ar@[parcolor]@{-}[ddr]
   & \neg a 
       \ar@[antiparcolor]@{--}[dd]
       \ar@[antiparcolor]@{--}[dr]
       \ar@[antiparcolor]@{--}[dll]
   & \\
b  \ar@[antiparcolor]@{--}[dr] \ar@[parcolor]@{-}[rrr]  
   &  &
   & \neg b 
     \ar@[parcolor]@{-}[dl]
   \\
   & c 
     \ar@[parcolor]@{-}[r]
   & \neg c &   
}}
\Rightarrow
\vcenter{
\xymatrix@R=14pt@C=14pt{
 a  \ar@[parcolor]@{-}[d] \ar@[parcolor]@{-}[dr]
    \ar@[antiparcolor]@{--}[r]      & 
 c  \ar@[parcolor]@{-}[d] \\
 \neg a \ar@[antiparcolor]@{--}[r] & \neg c
}
}
{\buildrel {\rm s.p.} \over \Rightarrow}
\vcenter{
\xymatrix@R=14pt@C=14pt{
 a  \ar@[parcolor]@{-}[d] \ar@[parcolor]@{-}[dr]
    \ar@[antiparcolor]@{--}[r]  & 
 c  \ar@[parcolor]@{-}[d] \ar@[parcolor]@{-}[dl] \\
 \neg a \ar@[antiparcolor]@{--}[r] & \neg c
}
}
\Rightarrow
\vcenter{
\xymatrix@R=14pt@C=14pt{
 \neg a  \ar@[parcolor]@{-}[d] \ar@[antiparcolor]@{--}[dr]
    \ar@[parcolor]@{-}[r] & 
 a  \ar@[parcolor]@{-}[d] \ar@[antiparcolor]@{--}[dl] \\
 c \ar@[antiparcolor]@{--}[r] & b
}
}.
$$
\item[5.] $b \parrel_T \neg c$. This case is analogous to case 2.
That is if we suppose that $\lnot(b \parrel_T \neg c)$, then the relation web:
$$
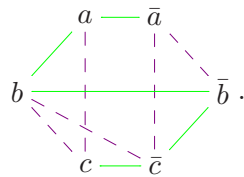
\vcenter{
\xymatrix@R=14pt@C=14pt{
   & a \ar@[parcolor]@{-}[dl]  \ar@[antiparcolor]@{--}[dd] 
       \ar@[parcolor]@{-}[r]
   & \neg a 
       \ar@[antiparcolor]@{--}[dd]
       \ar@[antiparcolor]@{--}[dr]
   & \\
 b  \ar@[antiparcolor]@{--}[dr] \ar@[parcolor]@{-}[rrr]  
    \ar@[antiparcolor]@{--}[drr]
   &  &
   & \neg b 
     \ar@[parcolor]@{-}[dl]
   \\
   & c 
     \ar@[parcolor]@{-}[r]
   & \neg c &   
}}.
$$
is isomorphic to (ii.1) in case 2.  
\item[6.] $\neg a \parrel_T c$. Suppose $\lnot(\neg a \parrel_T c)$. Since $a \parrel_T c$,
   as shown in Case 3, we have the following forbidden configuration:
$$
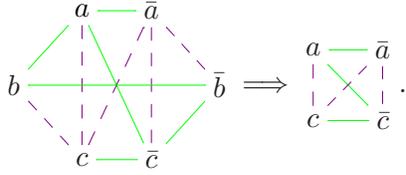
\vcenter{
\xymatrix@R=14pt@C=14pt{
   & a \ar@[parcolor]@{-}[dl]  \ar@[antiparcolor]@{--}[dd] 
       \ar@[parcolor]@{-}[r]
       \ar@[parcolor]@{-}[ddr]
   & \neg a 
       \ar@[antiparcolor]@{--}[dd]
       \ar@[antiparcolor]@{--}[dr]
       \ar@[antiparcolor]@{--}[ddl]
   & \\
b  \ar@[antiparcolor]@{--}[dr] \ar@[parcolor]@{-}[rrr]  
   &  &
   & \neg b 
     \ar@[parcolor]@{-}[dl]
   \\
   & c 
     \ar@[parcolor]@{-}[r]
   & \neg c &   
}}
\Longrightarrow
\vcenter{
\xymatrix@R=14pt@C=14pt{
 a  \ar@[antiparcolor]@{--}[d] \ar@[parcolor]@{-}[dr]
    \ar@[parcolor]@{-}[r] & 
 \neg a  \ar@[antiparcolor]@{--}[d] \ar@[antiparcolor]@{--}[dl] \\
 c \ar@[parcolor]@{-}[r] &  \neg c
}
}.
$$
\end{description}

We are left with the relations $a \seqrel_R c$, $b \seqrel_R c$, $\neg a \seqrel_R \neg b$
and $\neg a \seqrel_R \neg c$. Since the structural relation $ \seqrel $ can change
only to $ \aparrel $, there are 16 possible configurations to consider. 
By applying the triangular property, it is easy to see that
$a \aparrel_T c$ iff $b \aparrel_T c$, and that 
$\neg a \aparrel_T \neg b$ iff $\neg a \aparrel_T \neg c$. This is the case because
otherwise we would have the following forbidden configurations:
$$
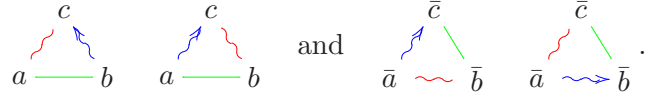
\vcenter{
\xymatrix@R=12pt@C=5pt{
 & c \ar@[aparcolor]@{~}[dl] & \cr
a \ar@[parcolor]@{-}[rr] &  & b \ar@[seqcolor]@{~>}[ul]
}}
\quad
\vcenter{
\xymatrix@R=12pt@C=5pt{
 & c \ar@[seqcolor]@{<~}[dl] & \cr
a \ar@[parcolor]@{-}[rr] &  & b \ar@[aparcolor]@{~}[ul]
}}
\quad
\hbox{{\rm and}}
\quad
\vcenter{\xymatrix@R=12pt@C=5pt{
 & \neg c \ar@[seqcolor]@{<~}[dl] & \cr
\neg a \ar@[aparcolor]@{~}[rr] &  & \neg b \ar@[parcolor]@{-}[ul]
}}
\quad
\vcenter{
\xymatrix@R=12pt@C=5pt{
 & \neg c \ar@[aparcolor]@{~}[dl] & \cr
\neg a \ar@[seqcolor]@{~>}[rr] &  & \neg b \ar@[parcolor]@{-}[ul]
}}.
$$
Therefore, we need only to consider the following three cases. 

\begin{description}
\item[7.] $a \aparrel_T c$, $b \aparrel_T c$, $\neg a \seqrel_T \neg b$ and 
          $\neg a \seqrel_T \neg c$.
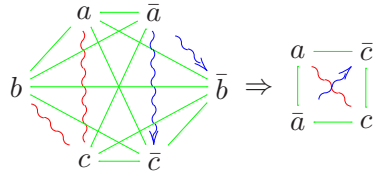
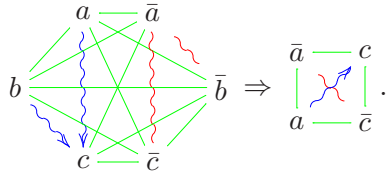
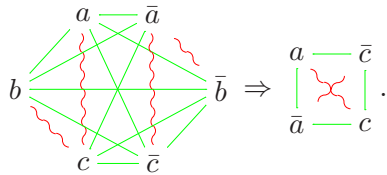
$$
\vcenter{
\xymatrix@R=14pt@C=14pt{
   & a \ar@[parcolor]@{-}[dl]  \ar@[aparcolor]@{~}[dd] \ar@[parcolor]@{-}[ddr]
       \ar@[parcolor]@{-}[drr] \ar@[parcolor]@{-}[r]
   & \neg a 
       \ar@[parcolor]@{-}[dll] \ar@[parcolor]@{-}[ddl] \ar@[seqcolor]@{~>}[dd]
       \ar@[seqcolor]@{~>}[dr]
   & \\
b  \ar@[aparcolor]@{~}[dr] \ar@[parcolor]@{-}[drr] \ar@[parcolor]@{-}[rrr]  
   &  &
   & \neg b 
     \ar@[parcolor]@{-}[dll]  \ar@[parcolor]@{-}[dl]
   \\
   & c 
     \ar@[parcolor]@{-}[r]
   & \neg c &   
}}
\Rightarrow
\vcenter{
\xymatrix@R=14pt@C=14pt{
 a \ar@[parcolor]@{-}[d] \ar@[aparcolor]@{~}[dr]
   \ar@[parcolor]@{-}[r] & 
 \neg c \ar@[parcolor]@{-}[d] \cr
 \neg a \ar@[parcolor]@{-}[r] 
        \ar@[seqcolor]@{~>}[ur] & c
}
}
$$ 
\item[8.] $a \seqrel_T c$, $b \seqrel_T c$, $\neg a \aparrel_T \neg b$ and 
          $\neg a \aparrel_T \neg c$.
$$
\vcenter{
\xymatrix@R=14pt@C=14pt{
   & a \ar@[parcolor]@{-}[dl]  \ar@[seqcolor]@{~>}[dd] \ar@[parcolor]@{-}[ddr]
       \ar@[parcolor]@{-}[drr] \ar@[parcolor]@{-}[r]
   & \neg a 
       \ar@[parcolor]@{-}[dll] \ar@[parcolor]@{-}[ddl] \ar@[aparcolor]@{~}[dd]
       \ar@[aparcolor]@{~}[dr]
   & \\
b  \ar@[seqcolor]@{~>}[dr] \ar@[parcolor]@{-}[drr] \ar@[parcolor]@{-}[rrr]  
   &  &
   & \neg b 
     \ar@[parcolor]@{-}[dll]  \ar@[parcolor]@{-}[dl]
   \\
   & c 
     \ar@[parcolor]@{-}[r]
   & \neg c &   
}}
\Rightarrow
\vcenter{
\xymatrix@R=14pt@C=14pt{
 \neg a \ar@[parcolor]@{-}[d] \ar@[aparcolor]@{~}[dr]
   \ar@[parcolor]@{-}[r] & 
 c \ar@[parcolor]@{-}[d] \cr
 a \ar@[parcolor]@{-}[r] 
        \ar@[seqcolor]@{~>}[ur] & \neg c
}
}.
$$ 
\item[9.] $a \aparrel_T c$, $b \aparrel_T c$, $\neg a \aparrel_T \neg b$ and 
          $\neg a \aparrel_T \neg c$.
$$
\vcenter{
\xymatrix@R=14pt@C=14pt{
   & a \ar@[parcolor]@{-}[dl]  \ar@[aparcolor]@{~}[dd] \ar@[parcolor]@{-}[ddr]
       \ar@[parcolor]@{-}[drr] \ar@[parcolor]@{-}[r]
   & \neg a 
       \ar@[parcolor]@{-}[dll] \ar@[parcolor]@{-}[ddl] \ar@[aparcolor]@{~}[dd]
       \ar@[aparcolor]@{~}[dr]
   & \\
b  \ar@[aparcolor]@{~}[dr] \ar@[parcolor]@{-}[drr] \ar@[parcolor]@{-}[rrr]  
   &  &
   & \neg b 
     \ar@[parcolor]@{-}[dll]  \ar@[parcolor]@{-}[dl]
   \\
   & c 
     \ar@[parcolor]@{-}[r]
   & \neg c &   
}}
\Rightarrow
\vcenter{
\xymatrix@R=14pt@C=14pt{
 a \ar@[parcolor]@{-}[d] \ar@[aparcolor]@{~}[dr]
   \ar@[parcolor]@{-}[r] & 
 \neg c \ar@[parcolor]@{-}[d] \cr
 \neg a \ar@[parcolor]@{-}[r] 
        \ar@[aparcolor]@{~}[ur] & c
}
}.
$$
\end{description}
The configurations in cases 7, 8 and 9 are forbidden by Lemma \ref{LemmaNoCross},
which means that the relations $a \seqrel_R c$, $b \seqrel_R c$, $\neg a \seqrel_R \neg b$
and $\neg a \seqrel_R \neg c$ must be preserved by $\rho$. Therefore, the configuration (i) 
must hold in $T$. 
\qed

\begin{lem}
\label{LemmaAlphaOne}
Let
$\vcenter{\ProofDia{\inf{\SizeZero{\rho}} {S\cons{P}} {U} }{}{\sysS}{d}{2} }$ be
a proof in a shallow system $\sysS$ where $\rho$ is a shallow rule, P is an $\alpha_n$-structure
and $S\cons{P}$ consists of pairwise distinct atoms. If $\sysS$ is equivalent to $\BV$
and all the structural relations in $\alpha_{n-1}$-substructures of $P$ are preserved by $\rho$
then all the structural relations in $P$ are also preserved by $\rho$.
\end{lem}
\proof
Let us recall the definition of $\alpha_n$-structure:
$$
P = \alpha_n(u,R,T) = 
      \pars{\seqs{\alpha_{n-1}(u.0, a_u, \pars{b_u, R});c_u}, 
            \seqs{\neg a_u;\alpha_{n-1}(u.1,\neg b_u, \pars{\neg c_u, T})}}.
$$
All the structural relations in $\alpha_{n-1}(u.0, a_u, \pars{b_u, R})$
(likewise, $\alpha_{n-1}(u.1,\neg b_u, \pars{\neg c_u, T})$) are preserved by $\rho$.

Let the variables $x,y$ stand for atoms in $\occ P$. The following case analyses
show that the structural relation between $x$ and $y$, for all possible values of
$x$ and $y$, must be preserved by $\rho$ as a consequence of the structural properties
of the underlying relation webs of $S\cons{P}$ and $U$.
We shall use the following abbreviation 
$$
P_1 = \alpha_{n-1}(u.0, a_u, \pars{b_u, R}), \hbox{ and }
P_2 = \alpha_{n-1}(u.1,\neg b_u, \pars{\neg c_u, T}),
$$
in the case analyses below.

\begin{description}
\item[1.] $x,y \in \occ P_1$, or $x,y \in \occ P_2$, follows immediately from
the condition of this lemma. 

\item[2.] $x \in  \occ P_1$, $y=c_u$. It holds in $P$:
          $x \seqrel y$.
\begin{description}          
\item[a.] $x \in \{a_u, b_u \}$. The structural relations between 
the atoms $a_u,$ $b_u,$ $c_u,$ $\neg a_u,$ $\neg b_u$ and $\neg c_u$  in $P$ are
$$
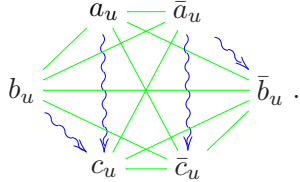
\vcenter{
\xymatrix@R=14pt@C=14pt{
   & a_u \ar@[parcolor]@{-}[dl]  \ar@[seqcolor]@{~>}[dd] \ar@[parcolor]@{-}[ddr]
       \ar@[parcolor]@{-}[drr] \ar@[parcolor]@{-}[r]
   & \neg a_u 
       \ar@[parcolor]@{-}[dll] \ar@[parcolor]@{-}[ddl] \ar@[seqcolor]@{~>}[dd]
       \ar@[seqcolor]@{~>}[dr]
   & \\
b_u  \ar@[seqcolor]@{~>}[dr] \ar@[parcolor]@{-}[drr] \ar@[parcolor]@{-}[rrr]  
   &  &
   & \neg b_u 
     \ar@[parcolor]@{-}[dll]  \ar@[parcolor]@{-}[dl]
   \\
   & c_u 
     \ar@[parcolor]@{-}[r]
   & \neg c_u &   
}
}.
$$
Note that since 
$a_u, b_u \in \occ P_1$ and $\neg b_u, \neg c_u \in \occ P_2$, 
and since by assumption all structral relations inside $P_1$ and $P_2$ are unchanged by $\rho$,
we have $a_u \parrel_U b_u$ and $\neg b_u \parrel_U \neg c_u$.
Therefore, by Lemma \ref{LemmaSnConfig}, the above configuration is preserved 
by $\rho$.

\item[b.] $x \not \in \{a_u, b_u \}$. Then $x \parrel_P a_u$ or $x \parrel_P b_u$,
     because by Definition~\ref{DefFractalStruct}
     $$\alpha_{n-1}(u.0, a_u, \pars{b_u, R}) = 
               \pars{Q_1\cons{a_u}, Q_2\cons{b_u}},$$ 
     for some structure contexts $Q_1\ec$ and $Q_2\ec$.
     Depending on whether $x \parrel_P b_u$ or $x \parrel_P c_u$, one of the following
     configurations must hold in $P$:
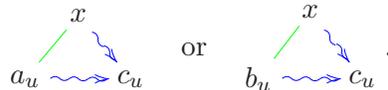
$$
\vcenter{
\xymatrix@R=12pt@C=5pt{
 & x \ar@[parcolor]@{-}[dl] & \cr
a_u \ar@[seqcolor]@{~>}[rr] &  & c_u \ar@[seqcolor]@{<~}[ul]
}}
\quad
\hbox{{\rm or}}
\quad
\vcenter{
\xymatrix@R=12pt@C=5pt{
 & x \ar@[parcolor]@{-}[dl] & \cr
b_u \ar@[seqcolor]@{~>}[rr] &  & c_u \ar@[seqcolor]@{<~}[ul]
}}.
$$
     The relations between $x$ and $\{ a_u, b_u \}$ are preserved by $\rho$ (they all belong to 
     $P_1$), so it holds in $U$: $x \parrel_U a_u$ or $x \parrel_U b_u$. The structural relations 
     $a_u \seqrel_U c_u$ and $b_u \seqrel_U c_u$ must hold as shown in Case 2.a. 
     The structural relation $x \seqrel_U c_u$ must hold in $U$, because otherwise
     $x \aparrel_U c_u$ by Definition~\ref{DefStructOrder} and one of the following forbidden 
     configuration would occur in $U$, depending on whether $x \parrel_U a_u$ or $x \parrel_U b_u$. 
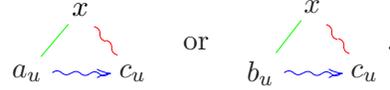
$$
\vcenter{
\xymatrix@R=12pt@C=5pt{
 & x \ar@[parcolor]@{-}[dl] & \cr
a_u \ar@[seqcolor]@{~>}[rr] &  & c_u \ar@[aparcolor]@{~}[ul]
}}
\quad
\hbox{{\rm or}}
\quad
\vcenter{
\xymatrix@R=12pt@C=5pt{
 & x \ar@[parcolor]@{-}[dl] & \cr
b_u \ar@[seqcolor]@{~>}[rr] &  & c_u \ar@[aparcolor]@{~}[ul]
}}.
$$
\end{description}

\item[3.] $x \in \occ P_2$, $y = \neg a_u$.
It holds in $P$: $y \seqrel_P x$.
\begin{description}
\item[a.] $x \in \{ \neg b_u, \neg c_u \}$, see Case 2.a.
\item[b.] $x \not \in \{ \neg b_u, \neg c_u \}$, then $x \parrel_U \neg b_u$ 
         or $x \parrel_U \neg c_u$.
         Using a similar argument as in Case 2.b, we establish that the structural relation 
         $\neg a_u \seqrel_U x$ must hold, because otherwise
         the following  forbidden configurations would occur in $U$:
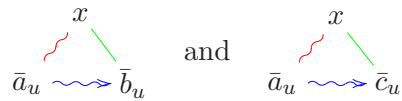
$$
\vcenter{
\xymatrix@R=12pt@C=5pt{
 & x \ar@[aparcolor]@{~}[dl] & \cr
\neg a_u \ar@[seqcolor]@{~>}[rr] &  & \neg b_u \ar@[parcolor]@{-}[ul]
}}
\quad
\hbox{{\rm and}}
\quad
\vcenter{
\xymatrix@R=12pt@C=5pt{
 & x \ar@[aparcolor]@{~}[dl] & \cr
\neg a_u \ar@[seqcolor]@{~>}[rr] &  & \neg c_u \ar@[parcolor]@{-}[ul]
}}
$$
which correspond to the cases $x \parrel_U \neg b_u$ and $x \parrel_U \neg c_u$, 
respectively.
\end{description}

\item[4.] $x \in \occ \seqs{P_1;c_u }$,
          $y \in \occ \seqs{\neg a_u; P_2}$.
\begin{description}          
\item[a.] $x \in \{ a_u, b_u, c_u \}$, $y \in \{ \neg a_u, \neg b_u, \neg c_u \}$, 
          see Case 2.a.
\item[b.] $x \in \{ a_u, b_u, c_u \}$, $y \not \in \{ \neg a_u, \neg b_u, \neg c_u \}$.
Since $y \in \occ P_2$, by assumption we have either $y \parrel_U \neg b_u$ or $y \parrel_U \neg c_u$.  
The strutural relations between atoms in $\{ x, \neg a_u,\neg b_u,\neg c_u \}$
are preserved by $\rho$, as shown in Case 2.a, and the structural relations between
atoms in $\{ y, \neg a_u,\neg b_u,\neg c_u \}$ are also preserved by $\rho$, as shown in 
Case 1 and Case 3. Therefore, if we change the relation $x \parrel_P y$ to 
$\lnot(x \parrel_U y)$, the square property will be violated, as shown in the following 
diagrams:
$$
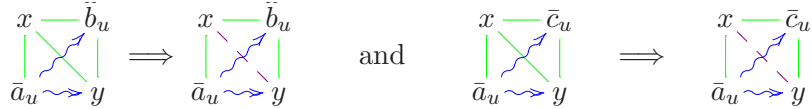
\vcenter{
\DiaSquareOneA{\neg a_u}{y}{\neg b_u}{x}
}
\Longrightarrow
\vcenter{
\DiaSquareOneB{\neg a_u}{y}{\neg b_u}{x}
}
\qquad
\hbox{{\rm and}}
\qquad
\vcenter{
\DiaSquareOneA{\neg a_u}{y}{\neg c_u}{x}
}
\quad
\Longrightarrow
\quad
\vcenter{
\DiaSquareOneB{\neg a_u}{y}{\neg c_u}{x}
}
$$
which correspond to the cases $y \parrel_U \neg b_u$ and $y \parrel_U \neg c_u$, respectively.
\item[c.] $x \not \in \{ a_u,b_u,c_u \}$, $y \in \{ \neg a_u, \neg b_u, \neg c_u \}$.
Then it holds in $U$: $x \parrel_U a_u$ or $x \parrel_U b_u$. 
The structural relations between atoms in $\{x,a_u,b_u, c_u \}$
must be preserved by $\rho$, as shown in Case 1 and 2, and the structural relations 
between atoms in $\{y, a_u, b_u, c_u \}$ must also be preserved as shown in Case 2.a.
The changing of $x \parrel_P y$ to $\lnot(x \parrel_U y)$ leads to the violation of square 
property as shown in the following diagrams:
$$
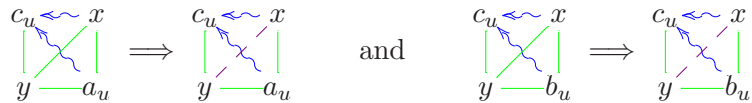
\vcenter{
\DiaSquareTwoA{y}{a_u}{x}{c_u}
}
\Longrightarrow
\vcenter{
\DiaSquareTwoB{y}{a_u}{x}{c_u}
}
\qquad
\hbox{{\rm and}}
\qquad
\vcenter{
\DiaSquareTwoA{y}{b_u}{x}{c_u}
}
\Longrightarrow
\vcenter{
\DiaSquareTwoB{y}{b_u}{x}{c_u}
}
$$
which correspond to the cases $x \parrel_U a_u$ and $x \parrel_U b_u$, respectively.
\item[d.] $x \not \in \{a_u,b_u,c_u \}$, $y \not \in \{\neg a_u,\neg b_u, \neg c_u \}$.
The structural relations $\neg a_u \parrel_P c_u$, $x \seqrel_P c_u$, $\neg a_u \seqrel_P y$,
$c_u \parrel_P y$ and $\neg a_u \parrel_P x$ are preserved by $\rho$
as shown in Case 3, 4.a, 4.b and 4.c. Therefore, if the relation $x \parrel_P y$ is changed
to $\lnot(x \parrel_U y)$, the square property would be violated.
$$
\vcenter{\DiaSquareThreeA{\neg a_u}{y}{x}{c_u}}
\Longrightarrow
\vcenter{\DiaSquareThreeB{\neg a_u}{y}{x}{c_u}}
$$
\end{description}
\end{description}
\qed

\begin{lem}
\label{LemmaAlphaTwo}
{Let 
$\vcenter{\ProofDia{\inf{\SizeZero{\rho}}{S\cons{P}}{U} }{}{\sysS}{d}{2.5}}$ be a proof in 
a shallow system $\sysS$, where $P$ is an $\alpha_n$-struture and $\rho$ is a shallow rule. 
If $\sysS$ is equivalent to $\BV$ and all the structural relations
in every $\alpha_0$-substructure of $P$ are preserved by $\rho$, then
all the structural relations in $P$ are also preserved by $\rho$.
}
\end{lem}
\proof
The case where $P = \alpha_0(u,R,T)$ is obvious. So suppose that
$$
P = \alpha_n(u,R,T) = \pars{\seqs{\alpha_{n-1}(u.0, a_u, \pars{b_u,R}) ;c_u}, 
       \seqs{\neg a_u;\alpha_{n-1}(u.1, \neg b_u, \pars{\neg c_u, T})}}
$$
By inductive hypothesis, the structural relations in $\alpha_{n-1}$ substructures
of $P$ are preserved by $\rho$. This means, by Lemma \ref{LemmaAlphaOne}, that all 
the structural relations in $P$ must also be preserved by $\rho$.
\qed

\begin{thm}
\label{TheoremDeepNesting}
{
No shallow system can be equivalent to $\BV$.
}
\end{thm}

\proof
Suppose that there exists a shallow system $\sysS$ which is equivalent to $\BV$.
Let $n$ be the depth of $\sysS$. 
Consider the  structure $S_{n+1}$. By Lemma~\ref{LemmaSnProvable}, $S_{n+1}$ is provable in $\BV$.
Since $\sysS$ is equivalent to $\BV$, there must exist also a proof 
$$\vcenter{\ProofDia{\inf{\SizeZero{\rho}}{S_{n+1}}{T} }{}
          {\sysS}{dd}{1.5}}.$$
We can assume without loss of generality that the instance of $\rho$ above is
non-trivial, i.e., $T \not = S_{n+1}$. 

The rule $\rho$ cannot be an interaction rule since by Lemma~\ref{LemmaNoDualStrInSn} 
there are no substructures of the form $\pars{R, \overline{R}}$ in $S_{n+1}$. Therefore it
must be a shallow rule with depth at most $n$. From Lemma~\ref{LemmaDepthOfAlphaZero}
all the $\alpha_0$-substructures in $S_{n+1}$ are at depth $2(n+1)$, so all the structural
relations in any $\alpha_0$-substructures of $S_{n+1}$ must be preserved by $\rho$.
But by Lemma~\ref{LemmaAlphaTwo}, this implies that all the structural relations
in $S_{n+1}$ must also be preserved by $\rho$, that is, $T = S_{n+1}$, contrary
to the assumption that the application of $\rho$ is non-trivial. Therefore, $S_{n+1}$
is not provable in $\sysS$ and consequently, $\sysS$ cannot be equivalent to $\BV$.
\qed

\section{Conclusion}
\label{sec:conc}

We have shown that deep inference is necessary for System $\BV$, that is, 
any restriction on the depth of the applicability of the rules in $\BV$
would result in a strictly less expressive logical system. 
This result extends to all conservative extensions of $\BV$, for instance,
the extension of $\BV$ with the exponential modalities of linear logic \cite{GugStr01}. 
This result also shows in particular that finding a simple sequent system for $\BV$
is challenging, to say the least. At the more conceptual level, we hope that it
gives sufficient evident that the calculus of structures, especially {\em deep-inference},
is a non-trivial departure from sequent calculus as a formalism for presenting
logics.

System $\BV$ is not the only logical system that incorporates both commutative
and non-commutative connectives. 
Retor\'e's Pomset logic \cite{RetINRIA99} achieves the same thing.
However, Pomset logic is based on proof-nets, and its sequentialisation 
is still an open question. 
Retor\'e uses a presentation of proof-nets called the {\em R\&B-directed cographs},
which are essentially relation webs where the par links are implicit except when they 
connect dual atoms. The structure $S_0$ defined in Section~\ref{sec:deep}, for example,
corresponds to the directed cographs:
$$
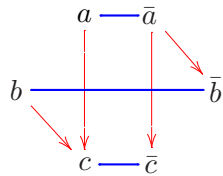
\xymatrix@R=14pt@C=14pt{
   & a \ar@[white]@{-}[dl]  \ar@[red]@{->}[dd] \ar@[white]@{-}[ddr]
       \ar@[white]@{-}[drr] \ar@[blue]@[thicker]@{-}[r]
   & \SmallSize{\neg a}
       \ar@[white]@{-}[dll] \ar@[white]@{-}[ddl] \ar@[red]@{->}[dd]
       \ar@[red]@{->}[dr]
   & \\
b  \ar@[red]@{->}[dr] \ar@[white]@{-}[drr] \ar@[blue]@[thicker]@{-}[rrr]  
   &  &
   & \SmallSize{\neg b}
     \ar@[white]@{-}[dll]  \ar@[white]@{-}[dl]
   \\
   & c 
     \ar@[blue]@[thicker]@{-}[r]
   & \SmallSize{\neg c} &   
}
$$
where the par links are omitted except those connecting dual atoms. 
The connection between $\BV$ and Pomset logic has been extensively discussed in \cite{StraThesis},
where it is shown a sound translation from $\BV$ to Pomset logic, i.e., every provable
structure (hence, relation webs) in $\BV$ corresponds to a proof net in Pomset logic.
Our result on the necessity of deep inference for $\BV$ therefore carries over to
Pomset logic. That is, certain types of sequent systems (i.e., those which can be
presented as shallow systems) can be ruled out in the search for the sequentialization
of Pomset logic.

\section*{Acknowledgments}
I would like to thank  Alessio Guglielmi, Kai Bruennler, Paola Bruscoli, Lutz Strassburger
and the anonymous referees for their useful comments and suggestions on earlier drafts
of this paper. An earlier version of this paper has appeared as a technical report \cite{Tiu01}
at TU Dresden.

\bibliographystyle{alpha}
\bibliography{biblio} 

\end{document}